\documentclass[prd,aps,preprint,showpacs,floats,nofootinbib,tightenlines]{revtex4}

\usepackage{epsfig}

\begin{document}
 
 \pagestyle{empty}
\preprint{
\noindent
\begin{minipage}[t]{3in}
\begin{flushright}
LBNL--3537E \\
\end{flushright}
\end{minipage}
}

\title{Approximate gauge symmetry of composite vector bosons}

\author{Mahiko Suzuki}
\affiliation{
Lawrence Berkeley National Laboratory and Department of Physics\\
University of California, Berkeley, California 94720
}

\date{\today}

\begin{abstract}

It can be shown in a solvable field theory model that the couplings of the 
composite vector bosons made of a fermion pair approach the gauge couplings 
in the limit of strong binding. Although this phenomenon may appear accidental 
and special to the vector bosons made of a fermion pair, we extend it to the 
case of bosons being constituents and find that the same phenomenon occurs 
in a more intriguing way. The functional formalism not only facilitates 
computation but also provides us with a better insight into the generating 
mechanism of approximate gauge symmetry, in particular, how the strong binding 
and global current conservation conspire to generate such an approximate 
symmetry. Remarks are made on its possible relevance or irrelevance to 
electroweak and higher symmetries.
 
\end{abstract}


\maketitle

\pagestyle{plain}

\setcounter{footnote}{0}

\section{Introduction}

Gauge symmetry is no doubt the underlying principle of contemporary particle 
theory. It is a mathematical or geometrical input rather than a dynamical 
consequence.  However, some people wonder if there is any physical or 
dynamical reason that necessitates gauge symmetry\cite{Heisenberg,Nielsen}.
Aside from such attempts, one suggestion was made 
decades ago\cite{Mandel} that when spin-one bosons are generated as tightly 
bound composite particles their couplings obey gauge invariance in the limit 
of vanishing mass. This was indeed demonstrated in a solvable model with 
fermions as constituents\cite{Suzuki}. As in most other attempts, the model 
was based on the Lagrangian of the Nambu-Jona-Lasinio type with a vector 
coupling and was, therefore, unrenormalizable, which is the price to pay for 
solvability. The conclusion was that all gauge-symmetry breakings as well as 
unrenormalizability are transformed into the mass of the composite spin-one 
bosons and that the composite boson mass can be made as small as one likes, 
but never zero, by making the binding force stronger, that is, infinitely 
close to gauge bosons but not exactly. Failure to realize genuine gauge 
invariance is obvious since the Lagrangian written in the fields of fermion 
constituents explicitly violates gauge invariance; perfect gauge symmetry 
should not arise where underlying dynamics explicitly violates it. 
Nonetheless, it is remarkable that gauge noninvariance is entirely 
transformed into the composite boson mass term.
   
The gauge boson sector of the electroweak model without a Higgs boson was 
built by the present author\cite{Suzuki} along this line twenty years ago
incorporating the proposal of Bjorken\cite{BJ} and of Hung and 
Sakurai\cite{HS}. Consistency of the large $N$ expansion as an effective 
low-energy field theory was analyzed for this model by Cohen, Georgi, 
and Simmons\cite{Georgi}, who also suggested how to incorporate quarks and 
leptons in this heretic electroweak model. It was immediately after 
production of the $W$ and $Z$ bosons were confirmed at CERN for the first 
time.  Since then, precision of the experimental measurement on the 
electroweak interaction has risen to test the standard model at the level 
of the loop corrections. Consequently, the phenomenological models of the 
late 1980's are no longer viable, but other options may still exist. Until 
we see an outcome of the Large Hadron Collider experiment, we should 
be prepared for possible surprises and leave all options open for 
phenomenology. It should be emphasized, however, that the purpose of the 
present paper is not to build a phenomenologically viable alternative to 
the standard electroweak theory, but to obtain a better understanding of 
the generation mechanism of approximate gauge invariance. Even if this 
mechanism may not turn out to be of use to model building in near future, 
it is an interesting theoretical subject of discussion in field theory. 

We shall find in this paper that the dynamical generation of approximate 
gauge symmetry is not an accident in the Nambu-Jona-Lasinio model or 
special to the fermionic constituents. 
A natural question arises as to how general this phenomenon is and 
which inputs are really necessary for this phenomenon to occur. The present 
paper first investigates the original fermionic constituent model by a 
different method and then moves on to explore how the approximate gauge 
symmetries are generated in the case of bosonic constituents, if at all. 
After studying the bosonic case, we understand the generation mechanism 
better and feel more confident that the mechanism is quite general and 
independent of specific models. 

   It may appear that our study has some technical resemblance with 
the phenomenon known as hidden symmetry, the name coined by 
Bando, Kugo, and Yamawaki\cite{Bando}. However, the hidden symmetry is 
something that is built in a theory at the beginning in one way or 
another. In contrast, we are concerned with the dynamics in which 
a relevant local symmetry does not exist, hidden or otherwise, at the 
fundamental level, but emerges only as an approximate symmetry in the 
low-energy effective Lagrangian.  In our case the local symmetry is 
explicitly broken at all levels.  We study how the explicit breakings 
of local symmetries transform into the Lagrangian of composite vector 
bosons. Our study focuses on a different subject, technically and 
conceptually, as we shall later comment more.

\section{Case of fermion constituents}

We start with a short summary of the results from an earlier 
paper\cite{Suzuki}. Let us think of forming tightly bound vector bosons 
out of fermions with heavy mass $M$. We choose that the fermions
transform like the fundamental representation of SU(n), which we refer 
to as  the ``flavor  group''.  The flavor group may be any other group.
In order to solve field theory explicitly, we choose the Lagrangian of 
the Nambu-Jona-Lasinio-type model with $N$ families of fermions and 
make the large $N$ expansion.

   In reasonably short-handed notations, the Lagrangian is written as
\begin{eqnarray}
 L(\overline{\psi},\psi) &=& \overline{\psi}(i/\!\!\!\partial - M)\psi 
                 - (G/2N)\sum_aj_{a\mu}j_a^{\mu}, \nonumber \\
  j_{a\mu} &=& \sum_{i=1}^N\overline{\psi}\gamma_{\mu}(\lambda_a/2)\psi.
\end{eqnarray} 
where $\frac{1}{2}\lambda_a$ ($a=1,2,3\cdots n^2-1$) are the generators of 
the flavor SU(n) in the $n\times n$ matrices, and the currents 
$j_{a\mu}$ ($a=1,2,3\cdots n^2-1$) are singlets of the U(N) family symmetry. 
The summation over flavor and family indices has been entirely suppressed 
in the kinetic energy and mass term.  While this Lagrangian is symmetric 
under the global SU(n)$\times$U(N) symmetry, it is obviously not invariant 
under gauge rotations of SU(n) or U(N) on $\psi/{\overline{\psi}}$. 
When the coupling constant 
$G$ is positive and larger than some critical value, the interaction 
generates vector bound states of a family singlet that form the adjoint 
representation of the flavor SU(n). In the leading $N$ order, explicit 
computation of the infinite fermion chain in Fig. 1 allows us to obtain for 
the bound states not only the mass and the coupling to the fermions but also 
the triple and quartic self-couplings\cite{Suzuki}. Although quadratic 
divergence does not appear in the loop diagrams thanks to global current 
conservation, logarithmic divergences do. We regularize them by the dimensional
regularization. The result is remarkable: All the couplings of the composite 
vector bosons obey SU(n) gauge invariance. The effective Lagrangian written 
in terms of $A_{\mu}$ and $\psi/\psi^{\dagger}$ reads in the standard notation 
\begin{equation}
L({\bf A}_{\mu},\overline{\psi},\psi) =
  -\frac{1}{2}{\rm tr}{\bf G}_{\mu\nu}{\bf G}^{\mu\nu}
  + m^2{\rm tr}{\bf A}_{\mu}{\bf A}^{\nu}
     +\overline{\psi}(i/\!\!\!\partial-g/\!\!\!\!{\bf A}-M)\psi, \label{L}
\end{equation} 
where ${\bf A}_{\mu}$ denotes the SU(n) adjoint vector fields in an 
$n\times n$ matrix and ${\bf G}_{\mu\nu}=\partial_{\mu}{\bf A}_{\nu}
-\partial_{\nu}{\bf A}_{\mu}+ig[{\bf A}_{\mu},{\bf A}_{\nu}]$.
The gauge coupling constant $g$ and the boson mass are obtained from the
loop diagram as
\begin{eqnarray}
      g^2 &=& 24\pi^2/N\ln(\overline{\Lambda}^2/M^2),\nonumber \\
      m^2 &=& 24\pi^2/G\ln(\overline{\Lambda}^2/M^2),   \label{gm}
\end{eqnarray}
where $\ln\overline{\Lambda}^2 \equiv (2-D/2)^{-1} +\ln 4\pi -\gamma_E$
in the dimensional regularization. It is only the mass term of the composite 
vector-boson fields ${\bf A}_{\mu}$ that is not gauge invariant in the 
effective Lagrangian of Eq. (\ref{L}). Furthermore, the four-fermion 
interactions of the original Lagrangian of the constituent fermions 
disappear from $L({\bf A}_{\mu},\overline{\psi},\psi)$, i.e., 
nonrenormalizability of the current-current interaction is also 
transferred entirely into this composite boson mass term. 

\noindent
\begin{figure}[h]
\epsfig{file=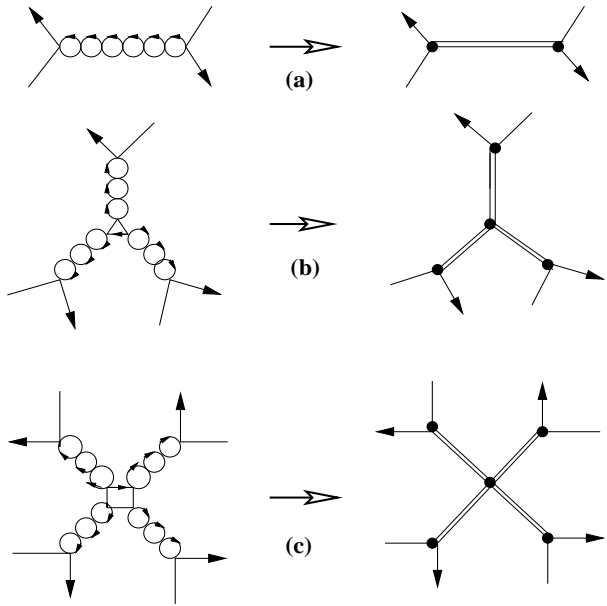,width=8cm,height=8cm} 
\caption{The infinite chain of fermion loops that generate
(a) a bound state and its coupling to the fermions, (b) the triple
self-coupling, and (c) the quartic self-coupling. 
The thin double lines denote bound states here and also in the figures
in the rest of the paper.
\label{fig:1}}
\end{figure}

   The composite boson mass squared $m^2$ can be made as small as one likes 
by increasing the magnitude of the coupling of binding $G$, but it can never 
reach zero for any finite value of $G$. This is because  each loop of 
the fermion self-energy chain is of the transverse form 
$(-g_{\mu\nu}+q_{\mu}q_{\nu}/q^2)\Pi(q^2)$ with $\Pi(0)=0$ by the global 
current conservation of SU(n) symmetry and consequently iteration of the 
loops leads to the fermion-antifermion scattering amplitude of the form, 
\begin{eqnarray}
 T  &=& -\biggl(\frac{G}{N}\biggr)\biggl(1-G\Pi(q^2)+\Big(G\Pi(q^2)\Bigr)^2
              - G\Big(G\Pi(q^2)\Bigr)^3 + \cdots \biggr), \nonumber \\
      &=& -\biggl(\frac{G}{N}\biggr)\frac{1}{1 + G\Pi(q^2)}.  \label{Pi}
\end{eqnarray}
Since $\Pi(q^2)\rightarrow \Pi'(0)q^2$ with $\Pi'(0)<0$ near $q^2=0$, 
the location of the bound-state pole determined by $1+G\Pi(q^2)=0$ is 
at $q^2\simeq 1/G|\Pi'(0)|$. Notice the importance of $\Pi(0)=0$ in order to 
have $m^2\sim 1/G$. This cannot be realized without global SU(n) 
current conservation. For bound states other than spin-parity $1^-$,
the natural scale of mass squared is $O(M^2)$ or else $O(\Lambda^2)$ 
($\Lambda$ = momentum cutoff). Furthermore we should appreciate that the 
same global current conservation prevents us from bringing the vector 
bound-state pole to $q^2=0$. This is consistent with the general 
theorem\cite{Case} by Case-Gasiorowicz and Weinberg-Witten that asserts 
incompatibility of the charged {\em massless} vector bosons with the 
Lorentz-covariant conserved currents carrying nonvanishing 
charges.  Putting it more simply, a massless spin-one boson cannot be 
obtained in the continuous limit of a massive spin-one boson, as we all 
know.  If we took literally the limit of $G\to\infty$, the mass $m^2$ 
would become zero, i.e., the composite bosons would look like gauge 
bosons. In this limit the entire Lagrangian would become $\propto 
-j_{a\mu}j_a^{\mu}$ alone after rescaling of the fields and therefore 
trivially gauge invariant.  However, the global currents would not exist 
by the Noether theorem in this pathological gauge-invariant limit. 
Conflict with the theorem could be thus evaded, but this limit is a case 
of no interest, physically or mathematically. 

How can these gauge-invariant couplings be generated ?  It is easy to 
understand when one works in the functional integral 
method\cite{Georgi,Bando}. The partition function $Z$ in terms of the 
$\psi/\overline{\psi}$ fields is given in the Euclidean metric by 
\begin{equation}
    Z = \int {\cal D}\overline{\psi}{\cal D}\psi
             \exp \int L(\overline{\psi},\psi) d^4x.
                            \label{Af}
\end{equation}  
We can replace the current-current interaction by introducing the auxiliary 
adjoint vector fields ${\bf A}_{\mu}=\sum_a(\lambda_a/2)A_{a\mu}$ as
\begin{equation}
    Z = \int {\cal D}{\bf A}_{\mu}{\cal D}\overline{\psi}{\cal D}\psi 
                  \exp \int \Bigl(L(\overline{\psi},\psi)
                + L_{{\rm aux}}({\bf A}_{\mu},\overline{\psi},\psi)\Bigr) d^4x,
                                      \label{AA}
\end{equation}
where the added Lagrangian term $L_{\rm aux}$ is defined in the Minkowski 
metric by
\begin{equation}
   L_{{\rm aux}}({\bf A}_{\mu},\overline{\psi},\psi) = 
   \frac{1}{2}\Bigl(mA_{a\mu}-\sqrt{G/N}j_{a\mu}\Bigr)
              \Bigl(mA_a^{\mu }-\sqrt{G/N}j_a^{\mu}\Bigr).  \label{LA}
\end{equation} 
Summation over the flavor $a$ is understood above and in the following.
Equivalence of the two actions in Eqs. (\ref{Af}) and (\ref{AA}) is 
obvious since one can trivially integrate out the fields ${\bf A}_{\mu}$ 
in Eq. (\ref{AA}) after shifting ${\bf A}_{\mu}$ in the functional space. 
When we open up $L_{\rm aux}({\bf A}_{\mu},\overline{\psi},\psi)$ and add 
it to $L(\overline{\psi},\psi)$, the current-current interactions cancel 
out between $L(\overline{\psi},\psi)$ and $L_{\rm aux}$, leaving the 
effective Lagrangian $L_{\rm eff}$ in the Minkowski metric in the form of 
\begin{eqnarray}
 L_{\rm eff}({\bf A}_{\mu},\overline{\psi},\psi) 
      &=& L(\overline{\psi},\psi)+L_{\rm aux}({\bf A}_{\mu},\overline{\psi},\psi), 
                              \nonumber \\
      &=& \overline{\psi}(i/\!\!\!\partial - M)\psi 
                 + \frac{1}{2}m^2 A_{a\mu}A_a^{\mu} 
                 - \sqrt{G/N}m j_{a\mu} A_a^{\mu}.   \label{LfA}
\end{eqnarray}
The constant in front of $j_{a\mu}A_a^{\mu}$ should be identified with the 
gauge coupling $g$ so that 
\begin{equation}
         g^2=(G/N)m^2. \label{constraint}
\end{equation}
The functional integration over ${\bf A}_{\mu}$ in Eq. (\ref{AA}) is equivalent 
to rewriting the current-current interaction with the ${\bf A}_{\mu}$ exchange 
at zero-momentum transfer. This explains the relation in Eq. (\ref{constraint}) 
as $\frac{1}{2!}g^2/m^2= G/2N$. While the diagram calculation has 
determined $g^2$ and $m^2$ individually as given in Eq. (\ref{gm}), the mass 
$m^2$ in Eq. (\ref{LfA}) is still a free parameter.  The reason is that
we have not yet incorporated the dynamical information of the fermion loop 
at this stage of the functional integral formulation.

The Lagrangian $L_{\rm eff}({\bf A}_{\mu},\overline{\psi},\psi)$ has no kinetic 
energy term of ${\bf A}_{\mu}$ so that its equation of motion for $A_{\mu}$ reads
\begin{equation}
   A_{a\mu} =(\sqrt{G/N}/m) j_{a\mu} = (g/m^2)
              \overline{\psi}\gamma_{\mu}(\lambda_a/2)\psi. \label{eq}
\end{equation}
It simply means that, before letting the vector-bosons propagate, they are 
made of fermion-antifermion pairs.  The global current conservation 
$\partial_{\mu}j_a^{\mu} = 0$ assures that the composite fields $A_{a\mu}$ 
consist only of spin-one states by $\partial_{\mu} A^{\mu}_a =(g/m^2)
\partial_{\mu}j_a^{\mu}=0$ leaving out the O(3) scalar component at this stage.

\noindent
\begin{figure}[h]
\epsfig{file=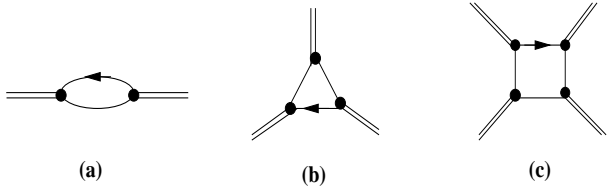,width=8cm,height=2.5cm} 
\caption{The diagrams that generate the kinetic energy term and the self-couplings
of the fields ${\bf A}_{\mu}$ in the case of fermionic constituents.    
\label{fig:2}}
\end{figure}

We can now proceed to generate the kinetic energy term and the self-couplings
of dimension four for ${\bf A}_{\mu}$ from the loop diagrams. 
With $L_{\rm eff}$ of Eq. (\ref{LfA}), iteration of the fermion loops no
longer occurs for the two-point function of ${\bf A}_{\mu}$ since 
there is no four-fermion interaction left in $L_{\rm eff}$. The relevant 
diagrams are only the single fermion-loop diagrams of ${\bf A}_{\mu}$ in 
the leading order of $N$. [Fig. 2(a)]. The same is true for the three and 
four-point functions. 
We should notice here that $L({\bf A}_{\mu},\overline{\psi},\psi)$ is gauge 
invariant up to the mass term of ${\bf A}_{\mu}$ since the interaction 
$-gj_{a\mu}A_a^{\mu}$ can be combined with the fermion kinetic energy 
term into the gauge-invariant form; 
\begin{equation}
  \overline{\psi}(i(/\!\!\!\partial+ig/\!\!\!\!{\bf A})-M)\psi.    
\end{equation}
The painstaking diagram calculation\cite{Suzuki} for the two, three and 
four-point functions of ${\bf A}_{\mu}$ was actually unnecessary; they 
must come out in the gauge-invariant combination up to the overall constant 
since the sole term of gauge noninvariance, namely, the vector-meson mass 
term $\frac{1}{2}m^2A_{a\mu}A_a^{\mu}$, does not enter the loop calculation
in the leading $N$ order. Therefore the radiatively produced Lagrangian of 
dimension four for the fields  ${\bf A}_{\mu}$ ought to be in the form
\begin{equation} 
   -Z_3\times\frac{1}{2}{\rm tr}{\bf G}_{\mu \nu}{\bf G}^{\mu\nu}.  
\end{equation}
An explicit loop-diagram calculation is needed only to obtain the
constant $Z_3$.  The fermion loop of Fig. 2a gives
\begin{equation}
   Z_3 = \frac{g^2N}{24\pi^2}\ln(\overline{\Lambda}^2/M^2).
    \label{Z3} 
\end{equation}
The constant $Z_3$ is absorbed into the wave-function renormalization of 
${\bf A}_{\mu}$ by ${\bf A}_{\mu} \rightarrow {\bf A}_{\mu}/\sqrt{Z_3}$,
which in turn renormalizes the coupling and the vector-boson mass too;
\begin{eqnarray}
    g &\rightarrow&  \sqrt{Z_3}g_r, \nonumber \\
    m^2 &\rightarrow&  Z_3 m^2_r.                  \label{renorm}
\end{eqnarray}
Because of $\Pi(0)=0$ there is no additive mass renormalization when the 
mass is computed at zero momentum. After the kinetic energy term of 
${\bf A}_{\mu}$ is computed and the renormalization of Eq. (\ref{renorm}) 
is performed, the renormalized coupling is given by
\begin{equation}
     g_r^2 = g^2/Z_3 =24\pi^2/N\ln(\overline{\Lambda}^2/M^2) 
\end{equation}
and the mass takes the form of
\begin{eqnarray}
     m_r^2 &=& m^2/Z_3 = (N/GZ_3)g^2, \nonumber \\ 
           &=& 24\pi^2/G\ln(\overline{\Lambda}^2/M^2).
\end{eqnarray}
These agree with the results of the loop-diagram iteration, 
Eq. (\ref{gm}).  Namely, the values for $g^2$ and $m^2$ that were 
obtained in the calculation of the infinite chain of loops actually 
incorporate the renormalization of Eq. (\ref{renorm}). The complete 
effective Lagrangian written in $\psi$, $\overline{\psi}$, and 
${\bf A}_{\mu}$ thus takes the SU(n) gauge-invariant form up to 
the boson mass term as given in Eq. (\ref{L}) with the understanding 
that the renormalization of Eq. (\ref{renorm}) has already been done 
for the mass and the coupling.

Before moving on, we summarize this section: In the explicitly solvable 
model of fermionic constituents a set of composite spin-one bosons behave 
exactly like gauge bosons in the small limit of the composite boson mass 
even though the fundamental Lagrangian is explicitly gauge noninvariant.  
Gauge noninvariance stays but solely in the mass term of the composite 
bosons. Since the origin of the boson mass is not spontaneous breaking 
of gauge symmetry, there is no asymmetric vacuum condensate of a scalar 
field, elementary nor composite. A hidden symmetry can be introduced in 
the fermionic model of the Nambu-Jona-Lasinio type, if one wishes, by
using its language, but it is always broken in this case. There is no 
unbroken phase of the hidden symmetry except for the pathological limit 
of $G\to\infty$\cite{Bando}. 

We have chosen the four-fermion binding force here in order to 
demonstrate all solutions explicitly in the cutoff field theory.  
Rather than going into a discussion of phenomenological relevance, 
we explore in the succeeding sections whether this remarkable 
phenomenon of dynamical gauge-symmetry generation is realized in other 
models or not, specifically, in the case that the constituents are 
spinless bosons. The option of bosons being fundamental particles is 
even more esoteric phenomenologically and sounds less attractive.  
Our purpose here is, however, to obtain a better understanding of this
generation mechanism of an approximate gauge symmetry from other models. 

\section{Case of bosonic constituents: abelian symmetry}

  We would like to see whether the gauge-symmetry generation of the 
preceding section works in the bosonic constituent models or not. We 
emphasize that we do not slip an unbroken local symmetry in our models to 
look for massless vector bosons as dynamical gauge boson modes. Such a
study was done in the $CP^{N-1}$ a few decades ago; a local symmetry 
is present at the beginning as redundancy when its Lagrangian is written 
in some form, then a composite massless vector boson is searched for.  
Instead we choose models in which there is no local symmetry to start 
with. We study whether the explicit breaking can be transformed 
into the mass term alone in the case of tightly bound vector bosons.
Unlike the fermionic model, to our knowledge, our bosonic models have 
never been studied in the literature.  They show us more clearly 
what realizes an approximate gauge symmetry as a consequence of 
compositeness.   
     
  Let us first study the case of an Abelian vector boson since it gives
us a good insight into the problem leaving out unnecessary complications. 
We form a neutral composite vector boson (a massive photon) with charged 
spinless bosons like $\pi^{\pm}$ having heavy mass $M$. We introduce $N$ 
families of the heavy $\pi^{\pm}$ for the large $N$ expansion. Our 
fundamental Lagrangian is written in the nonpolynomial form as
\begin{eqnarray}
    L(\phi^{\dagger},\phi) 
    &=& \partial_{\mu}\phi^{\dagger}\partial^{\mu}\phi -M^2\phi^{\dagger}\phi
   -\biggl(\frac{G}{2N}\biggr)\frac{j_{\mu}j^{\mu}}{1+2(G/N)\phi^{\dagger}\phi}, 
                          \nonumber \\
   j_{\mu} &=& i \sum_i 
        \phi^{\dagger(i)}\stackrel{\leftrightarrow}{\partial}_{\mu}\phi^{(i)},
            \label{Lb}  
\end{eqnarray}  
with the charged spinless bosons $\phi^{(i)}$ and $\phi^{(i)\dagger}$ 
($i= 1,2,3,\cdots N$) of $N$ families.
No constraint is imposed on the fields $\phi$ and the classical vacuum is
at $\langle\phi\rangle = 0$ so that the Lagrangian is invariant under the 
global U(1) charge rotation,
\begin{equation}
   \phi(x)\rightarrow e^{i\alpha}\phi(x), \;\;
  \phi^{\dagger}(x)\rightarrow e^{-i\alpha}\phi^{\dagger}(x), 
\end{equation}
and trivially invariant under global U(N) family rotations. The 
current-current interaction has been so chosen that not only a tightly 
bound state can be formed but also its mass is explicitly calculable in 
the large $N$ limit.\footnote{It may look that the factor of 
$1/[1+2(G/N)\phi^{\dagger}\phi]$ in this Lagrangian has some vague 
resemblance with that of the $CP^{N-1}$ model written in the constrained 
fields\cite{Haber}. But our $\phi/\phi^{\dagger}$ are unconstrained here.}
 
A natural extension of the fermionic model might suggest the current-current 
interaction $(G/2N)j_{\mu}j^{\mu}$ of $\phi/\phi^{\dagger}$ in the bosonic
case. However, this simple current-current interaction does not generate 
a tightly bound vector boson in the scattering amplitude for the following 
reasons: 

1.  The current 
$j_{\mu}=i\phi^{\dagger}\stackrel{\leftrightarrow}{\partial}_{\mu}\phi$,
is \underline{not} a conserved current in the case that the interaction 
is $-(G/2N)j_{\mu}j^{\mu}$. Because derivatives of $\phi/\phi^{\dagger}$ 
enter $-(G/2N)j_{\mu}j^{\mu}$, the conserved current\footnote{
Hereafter we denote the Noether currents with the capital letters and 
distinguish them from the naive bosonic currents 
$j_{\mu}=i\phi^{\dagger}\stackrel{\leftrightarrow}{\partial}_{\mu}\phi$ 
that originates from the kinetic energy term alone.}
$J_{\mu}$ derivable by the Noether theorem in the Abelian case,
\begin{equation}
 J_{\mu} =-i\biggl(\frac{\partial L}{\partial(\partial^{\mu}\phi)}\phi
    -\frac{\partial L}{\partial(\partial^{\mu}\phi^{\dagger})}\phi^{\dagger}
       \biggr)            \label{jN}
\end{equation}
contains a term which depends on the interaction $(G/2N)j_{\mu}j^{\mu}$.
If we went ahead with this naive current-current interaction
$L_{\rm int}=-(G/2N)j_{\mu}j^{\mu}$, the current $j_{\mu}$ would not 
conserve, $\partial^{\mu}j_{\mu}\neq 0$.  Its immediate consequence is 
that the self-energy loop $\Pi_{\mu\nu}(q)$ is not transverse 
($q^{\mu}\Pi(q)_{\mu\nu}\neq 0$) and an additional term of quadratic
divergence $O(\Lambda^2)$ arises with the coefficient $g_{\mu\nu}$. 
This moves the pole to $q^2 \sim 1/G - O(\Lambda^2)$ that cannot be 
physically interpreted as mass square of a bound state. 

2. From the standpoint of the functional formalism, choosing the simple 
interaction $-(G/2N)j_{\mu}j^{\mu}$ would amount to postulating that the 
composite vector field $A_{\mu}$ be proportional to $j_{\mu}$ and consequently 
lead to $\partial_{\mu}A^{\mu}\propto\partial_{\mu}j^{\mu}\neq 0$. That is, 
$A_{\mu}$ would not be purely a field of spin-one, but contain a spin-zero 
component.

We were fortunate in the model of fermionic constituents since the binding 
interaction contains no derivative of fields and therefore the choice of the 
interaction was deceptively simple. In contrast, for the bosonic constituents 
we must choose the interaction carefully such that the auxiliary composite 
field $A_{\mu}$ is proportional to the Noether current. If so chosen, the 
field $A_{\mu}$ obeys $\partial^{\mu}A_{\mu}=0$ and its proper self-energy 
part $\Pi_{\mu\nu}(q)$ turns out to be transverse. Only in this situation 
can the composite boson mass be made as small as one likes by increasing 
the binding interaction constant $G$. The factor $[1+2(G/N)\phi^{\dagger}\phi]$ 
in the denominator of the interaction in Eq. (\ref{Lb}) serves this purpose 
and realizes $m^2\propto 1/G\rightarrow 0$. [See the second relation in 
Eq. (\ref{gm}).]

\subsection{Diagram computation}

    A neutral vector bound-state is formed with the loop and bubble diagrams
of $\phi/\phi^{\dagger}$. We compute for the bound-state in elastic scattering
\begin{equation}
     \phi^+(p_1) + \phi^-(p_2) \rightarrow \phi^+(p_3) + \phi^-(p_4).
                            \label{scattering}
\end{equation} 
Although we are interested in physics at large $G$, we cannot make the $1/G$
expansion in the Lagrangian since the potential term behaves at large $G$ as
\begin{equation}
 L_{\rm int}=-\biggl(\frac{j_{\mu}j^{\mu}}{4(\phi^{\dagger}\phi)}\biggr) 
       \biggl[1- \frac{N}{2G(\phi^{\dagger}\phi)}+\cdots\biggr].  
\end{equation}
The behavior of $L_{\rm int}\to \infty$ at $\phi^{\dagger}\phi=0$ makes
the perturbative vacuum ill-defined and prevents diagram calculation. We must
instead perform diagram calculation in the perturbative expansion in powers
of $G$ to all orders, sum up the perturbative series and then take $G$ to 
large values. Such computation is possible only in the leading $N$ order.
In the large $N$ limit the relevant diagrams are chains of loops with
bubbles added. (See Fig. 3.) Each loop comes from the diagram 
of $\langle 0|T(\phi(x)\phi^{\dagger}(y))|\rangle
\langle 0|T(\phi(y)\phi^{\dagger}(x))|0\rangle$, while the bubble diagram
arise from $\langle 0|T(\phi(x)\phi^{\dagger}(x))|0\rangle$ in the power 
series expansion of the factor $1/[1+2(G/N)\phi^{\dagger}(x)\phi(x)]$.
If the loops and the bubbles of $O(G^n)$ in the $J^P=1^-$ channel sum 
into the transverse form as 
\begin{equation}
     (-g_{\mu\nu} + q_{\mu}q_{\nu}/q^2)\Pi(q^2);\;\; \Pi(0) = 0,          
                    \;\;  (q=p_1+p_2), \label{SE}
\end{equation} 
the perturbation series turns into the total scattering amplitude in the 
form of Eq. (\ref{Pi}) so that the bound-state mass $m^2$ comes out to be 
$\propto 1/G$. Transversality of Eq. (\ref{SE}) is indeed realized after summing 
the loops and the bubbles of the same order in the power of $G$, as shown in Fig. 3.  
In $O(G)$ the tree diagram is the only diagram [Fig. 3(a)]. A single-loop 
diagram and a single-bubble diagram enter $O(G^2)$ [Fig. 3(b)]. The single-loop 
diagram alone would not make $\Pi_{\mu\nu}(q)$ transverse in $O(G^2)$, as we 
know from the photon self-energy in electrodynamics of the charged pions in  
which the bubble generated by the interaction 
$e^2\phi^{\dagger}\phi A_{\mu}A^{\mu}$ makes the photon self-energy 
transverse and keeps the photon massless even after loop corrections. 
In $O(G^3)$ we sum a diagram with two bubbles, a pair of diagrams with 
one-loop and one-bubble, and the diagram of two bubbles [Fig. 3(c)].  
We can keep on going to higher orders of $G$ and obtain a scattering amplitude 
of the form of Eq. (\ref{SE}) in the $1^-$ channel. Consequently, the location 
of the pole is found at $q^2\sim 1/G$ as we desire. In order to realize this 
behavior, therefore, the factor $1/[1+2(G/N)\phi^{\dagger}\phi]$ is 
needed in the interaction term of the Lagrangian, Eq. (\ref{Lb}).

\noindent 
\begin{figure}[h]
\epsfig{file=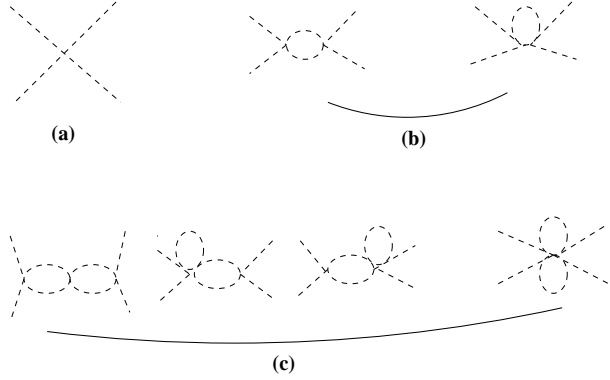,width=8cm,height=5cm} 
\caption{The diagrams in each order of $O(G^n)$ that sum up into the
transverse form and therefore generate a tightly bound vector-boson. 
 (a) $O(G)$, (b) $O(G^2)$, and (c) $O(G^3)$.     
\label{fig:3}}
\end{figure}

     When we sum up an infinite series of the loop plus bubble diagrams of
Fig. 3 in a compact form, the resulting invariant amplitude for the elastic 
scattering of Eq. (\ref{scattering}) is 
\begin{eqnarray}
 T &=& (G/N)g_{\mu\nu}(p_3-p_4)^{\mu}(p_1-p_2)^{\nu}
  [1 - G\Pi(q^2) +(-G\Pi(q^2))^2 + \cdots ],
  \nonumber  \\  
   &=& \biggl(\frac{G}{N}\biggr)\frac{(p_3-p_4)_{\mu}
               (p_1-p_2)^{\mu}}{1+G\Pi(q^2)},
\end{eqnarray}
where
\begin{equation}
    \Pi(q^2) = -\frac{q^2}{48\pi^2}\Bigl[\ln\overline{\Lambda}^2
  -3\int_0^1(1-2\alpha)^2\ln\Bigl(M^2-\alpha(1-\alpha)q^2\Bigr)d\alpha\Bigr].
\end{equation}
The bound-state pole appears in $T$ at the location determined by 
$1+G\Pi(q^2)=0$. For large $G$ the mass of the bound state is found at 
\begin{equation}
     m^2 = \frac{48\pi^2}{G\ln(\overline{\Lambda}^2/M^2)}, \label{m}
\end{equation} 
and its coupling to $\phi^{\pm}$ is given by
\begin{equation}
     g^2 = \frac{48\pi^2}{N\ln(\overline{\Lambda}^2/M^2)}, \label{g}
\end{equation}
so that the bound-state mass and the coupling are related at $GM^2\gg 1$ by  
\begin{equation}
     g^2 = (G/N)m^2.       \label {gmb}
\end{equation}
Here again the composite mass vanishes if one takes the limit of $G\to\infty$
in Eq. (\ref{m}). Going back to the original Lagrangian, Eq. (\ref{Lb}), we 
find in this limit\footnote{This limiting Lagrangian appeared in 
Reference \cite{Akh} in a different context.}
\begin{equation}
  \lim_{G\to\infty}  L(\phi^{\dagger},\phi) 
        = \partial_{\mu}\phi^{\dagger}\partial^{\mu}\phi -M^2\phi^{\dagger}\phi
           - j_{\mu}j^{\mu}/4(\phi^{\dagger}\phi).  \label{Llim}
\end{equation}
It is not difficult to check that this Lagrangian is U(1)-gauge invariant 
under $\phi(x)/\phi^{\dagger}(x)\to e^{i\chi(x)}\phi(x)/e^{-i\chi(x)}\phi^{\dagger}(x)$.
The U(1)-gauge invariance is nontrivial in this form unlike that in the fermionic model.  
In the nonlinear representation, however, this limiting Lagrangian turns out to be the
free Lagrangian of the radial fields and the phase fields do not enter. That is, this
Lagrangian is trivially gauge invariant and of little physical content.
The Noether current of Eq. (\ref{jN}) vanishes 
\begin{equation}
         J_{\mu} =0  \;\; (G\to\infty)
\end{equation}
Since there is not a conserved global U(1) current, the massless limit of our composite 
boson would not contradict the general theorem\cite{Case}. But it is obvious that
a massless vector boson cannot be formed with the limiting Lagrangian that contains 
only the Hermitian radial field.

The Lagrangian of Eq. (\ref{Llim}) would be identical to that of the $CP^{n-1}$
model \underline{if} $\phi/\phi^{\dagger}$ were constrained with $\phi^{\dagger}\phi 
= {\rm const}$.  However, the $\phi/\phi^{\dagger}$ fields are the unconstrained 
fields in our case; we have gone through the Feynman diagram calculation with 
the standard (unconstrained) spinless-boson propagator. 

\subsection{Functional integral formulation}

  In the diagram computation above, we need careful bookkeeping in summing 
up the perturbation series into the scattering amplitude.  After our study of 
the fermionic model, however, we are able to carry out an equivalent calculation 
by the functional integral method in a simpler way. We see underlying 
issues and their new aspects more clearly in a new light.

  The first step is to introduce the auxiliary neutral vector field $A_{\mu}$.
The Lagrangian of Eq. (\ref{Lb}) in $\phi/\phi^{\dagger}$ suggests the form
for the partition function,
\begin{equation}
    Z = \int {\cal D}A_{\mu}{\cal D}\phi^{\dagger}{\cal D}\phi\exp\Bigl[\int d^4x 
   \Bigl[L(\phi^{\dagger},\phi)+ L_{\rm aux}(A_{\mu},\phi^{\dagger},\phi)\Bigr], 
                      \label{AbA}  
\end{equation}
where $L(\phi^{\dagger},\phi)$ is given by Eq. (\ref{Lb}) and 
$L_{\rm aux}(A_{\mu},\phi^{\dagger}, \phi)$ is defined by 
\begin{eqnarray}
 L_{\rm aux}(A_{\mu},\phi^{\dagger},\phi) 
         &=& \frac{1}{2}\Bigl(1+2(G/N)\phi^{\dagger}\phi\Bigr)
    \biggl(mA_{\mu}-\frac{\sqrt{G/N}j_{\mu}}{1+2(G/N)\phi^{\dagger}\phi}\biggr)
    \biggl(mA^{\mu}-\frac{\sqrt{G/N}j^{\mu}}{1+2(G/N)\phi^{\dagger}\phi}\biggr)
                                  \nonumber \\
         &+& 2\delta^4(0)\ln\Bigl(1+2(G/N)\phi^{\dagger}\phi\Bigr).
             \label{Lbaux}
\end{eqnarray}
Upon integration over $A_{\mu}$, the factor $[1+2(G/N)\phi^{\dagger}\phi]$ 
in front of the first term of Eq. (\ref{Lbaux}) generates 
$\exp[-2\delta^4(0)\int\ln(1+2(G/N)\phi^{\dagger}\phi)d^4x]$, as is shown in the 
Appendix, and cancels the last term of $L_{\rm aux}$ so that Eq. (\ref{AbA}) 
reduces to the partition function written in $\phi/\overline{\phi}$ alone,
\begin{equation}
   Z = \int{\cal D}\phi^{\dagger}{\cal D}\phi\exp \int 
   L(\phi^{\dagger},\phi)d^4x.
\end{equation}
The diagrammatic content of this logarithmic term is also shown in the 
Appendix. The infinite factor $\delta^4(0)$ represents the total functional
phase space $\int d^4k/(2\pi)^4$, which should be properly regularized.
Physically, it is a large finite number since the unrenormalizable 
Lagrangian of Eq. (\ref{Lb}) is valid only up to some limited energy range.
However, if one regularizes it dimensionally as the $M^2\to 0$ limit of 
\begin{equation}
  \int\frac{d^Dk}{(2\pi)^D}\frac{k^2}{k^2-M^2} =
      \frac{i}{16\pi^2}M^4\biggl(\ln\frac{\overline{\Lambda}^2}{M^2}+1\biggr),
\end{equation}  
one would set this $\delta^4(0)$ to zero using the formula
\begin{equation}
  \int\frac{d^Dk}{(2\pi)^D}\frac{k^2}{(k^2-M^2)^N} =
         \frac{i}{2}\frac{(-1)^{N-1}}{(4\pi)^{D/2}}
          \frac{\Gamma(N-1-D/2)}{\Gamma(N)}\frac{D}{(M^2)^{N-1-D/2}}.
\end{equation}  
Containing no derivative, the term $2\delta^4(0)\ln(1+2(G/N)\phi^{\dagger}\phi)$ 
is manifestly gauge invariant by itself and, in the leading $N$ order, does not 
contribute to the calculation of the bound state in the $J^P=1^-$ channel. It affects 
only the $0^+$ channel of $\phi\phi^{\dagger}$ scattering in the leading $N$ order.
We leave this singular term as proportional to $\delta^4(0)$ as it is, while it 
does not affect our diagrammatic calculation in the rest of the paper. 
   
When we sum $L(\phi^{\dagger},\phi)$ and $L_{\rm aux}(A_{\mu},\phi^{\dagger},\phi)$,
no current-current interaction is left in the sum,
\begin{eqnarray}
  L(\phi^{\dagger},\phi)+ L_{\rm aux}(A_{\mu},\phi^{\dagger},\phi) &=&
   \partial_{\mu}\phi^{\dagger}\partial^{\mu}\phi -M^2\phi^{\dagger}\phi  
       +2\delta^4(0)\ln\Bigl(1+2(G/N)\phi^{\dagger}\phi\Bigr)
                    \nonumber \\
         &+& \frac{1}{2}m^2\Bigl(1+2(G/N)\phi^{\dagger}\phi\Bigr)
         A_{\mu}A^{\mu} - \sqrt{G/N}mj_{\mu}A^{\mu}. 
   \label{LphiA} 
\end{eqnarray}
In fact, we have chosen $L_{\rm aux}$ in Eq. (\ref{Lbaux}) so that the 
current-current interaction is absent from the sum in Eq. (\ref{LphiA}).
We have not obtained Eq. (\ref{LphiA}) by simply gauging $L(\phi^{\dagger},\phi)$ 
with $\partial_{\mu}\to D_{\mu}$ in Eq. (\ref{Lb}).
We identify the constant $\sqrt{G/N}m$ in front of $j_{\mu}A^{\mu}$ with the 
gauge coupling $g$ in Eq. (\ref{LphiA}). Therefore we obtain $g^2 = (G/N)m^2$, 
which is the relation between $g^2$ and $m^2$ that has been obtained 
in Eq. (\ref{gmb}) by the diagram calculation. Furthermore a new four-point 
interaction of $\phi^{\dagger}\phi A_{\mu}A^{\mu}$ arises 
in $L(\phi^{\dagger},\phi)+L_{\rm aux}(A_{\mu},\phi^{\dagger},\phi)$,
\begin{equation}
      (m^2 G/N)\phi^{\dagger}\phi A_{\mu}A^{\mu},
\end{equation}
which is equal to $g^2\phi^{\dagger}\phi A_{\mu}A^{\mu}$ thanks to 
$g^2 = (G/N)m^2$.  Therefore the Lagrangian in Eq. (\ref{LphiA}) can be 
rearranged into the gauge-invariant form up to the mass term of $A_{\mu}$:
\begin{eqnarray}
      L_{\rm eff}(A_{\mu},\phi^{\dagger},\phi) &\equiv&   
        L(\phi^{\dagger},\phi) + L_{\rm aux}(A_{\mu},\phi^{\dagger},\phi) \nonumber \\
      & = & (\partial_{\mu}\!-\!igA_{\mu})\phi^{\dagger}
              (\partial^{\mu}\!+\!igA^{\mu})\phi 
           +\!\frac{m^2}{2} A_{\mu}A^{\mu}
     +\! 2\delta^4(0)\ln\Bigl(\!1\!+\!2(G/N)\phi^{\dagger}\phi\Bigr).
                                      \label{Lg}             
\end{eqnarray}
The Lagrangian of Eq. (\ref{Lg}) leads to the equation of motion for $A_{\mu}$,
\begin{equation}
      m^2\Bigl(1+(2G/N)\phi^{\dagger}\phi\Bigr)A_{\mu}-\sqrt{G/N}mj_{\mu} = 0.
                          \label{eqofmotion} 
\end{equation}
The Noether current $J_{\mu}$ in terms of $\phi$ and $\phi^{\dagger}$ 
can be computed with Eq. (\ref{jN}) from $L(\phi^{\dagger},\phi)$ as
\begin{equation} 
   J_{\mu}= i\frac{\phi^{\dagger}\partial_{\mu}\phi-
      \partial_{\mu}\phi^{\dagger}\phi}{1+2(G/N)\phi^{\dagger}\phi}.
\end{equation}
Therefore, the equation of motion, Eq. (\ref{eqofmotion}), together with
$g^2=(G/N)m^2$ says that, with our choice of Lagrangian, $A_{\mu}$ is 
proportional to the Noether current $J_{\mu}$ of the constituent fields
before propagation;
\begin{equation}
      A_{\mu} = \frac{g}{m^2}J_{\mu}.
\end{equation}
Consequently it satisfies $\partial_{\mu}A^{\mu}=0$ so that its self-energy is
transverse.  Therefore, the bound-state mass square can behave as $m^2\sim 1/G$.  
The relation $\partial_{\mu}A^{\mu}=0$ also tells that this vector boson is 
not a gauge boson since it holds by the equation of motion, not by choice of 
fixing ambiguities.  Nor does $A_{\mu}$ transform like 
$A_{\mu}\to A_{\mu}+(i/g)\partial_{\mu}\chi$ under $\phi/\phi^{\dagger}
\to e^{i\chi}\phi/e^{-i\chi}\phi^{\dagger}$ either. That is, we are studying 
something very different from the gauge boson of the $CP^{N-1}$ 
model\cite{Haber} or its hidden symmetry\cite{Bando}.   

We now proceed to obtain the kinetic energy term through loop diagrams. 
No three-point function or nonderivative four-point function of $A_{\mu}$ 
is generated in the Abelian case. It is only the two-point functions that
arise from loop and bubble diagrams. The computation is straightforward by 
the diagrams of Fig. 4 with the interaction 
$-gj_{\mu}A^{\mu} +g^2\phi^{\dagger}\phi A_{\mu}A^{\mu}$.  

\noindent
\begin{figure}
\epsfig{file=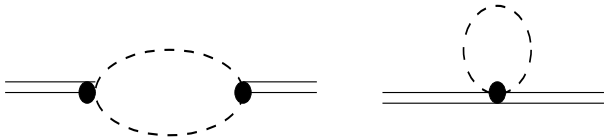,width=8cm,height= 1.8cm} 
\caption{The diagrams that generate the kinetic energy term of the composite 
boson field. No further chain of loops and bubbles enter in the leading $N$ order.     
\label{fig:4}}
\end{figure}

Since the interaction $-gj_{\mu}A^{\mu}+g^2A_{\mu}A^{\mu}\phi^{\dagger}\phi$ 
added with $\partial_{\mu}\phi^{\dagger}\partial^{\mu}\phi$ is gauge invariant 
and since the mass term of $A_{\mu}$ enters nowhere in this loop and bubble 
calculation, the resulting kinetic energy is also gauge invariant:
\begin{equation}
    -\frac{Z_3}{4}(\partial_{\mu}A_{\nu}-\partial_{\nu}A_{\mu})
              (\partial^{\mu}A^{\nu}-\partial^{\nu}A^{\mu}),
\end{equation}  
where the constant $Z_3$ is computed with the loop diagram from the Lagrangian 
of Eq. (\ref{Lg});
\begin{equation}
    Z_3 = \frac{g^2}{48\pi^2}\ln\frac{\overline{\Lambda^2}}{M^2}.
\end{equation}
This constant $Z_3$ is removed by the wave-function renormalization of the 
field $A_{\mu}$ and renormalization of the coupling $g$, and the mass $m^2$:
\begin{eqnarray}
   A_{\mu} &\rightarrow& A_{\mu}/\sqrt{Z_3} \nonumber \\
   g &\rightarrow& \sqrt{Z_3}g \nonumber \\
   m^2 &\rightarrow& Z_3 m^2.    
\end{eqnarray}
The renormalized mass and coupling are what the diagram computation has
given in Eqs. (\ref{m}) and (\ref{g}) as in the fermionic model. 
Even after they are renormalized, they maintain the relation $g^2 = (G/N)m^2$ 
of Eq. (\ref{gmb}).  We have thus confirmed the results of our preceding 
diagram calculation in the bosonic model and have reaffirmed our finding 
in the fermionic model:
In the small limit of the composite boson mass the bosonic theory also 
approaches the gauge theory as closely as possible although the mass $m^2$ 
can never be brought to zero for any large but finite value of $G$.

We have a little deeper understanding of the relation between gauge invariance 
and the small composite boson mass in the bosonic constituent model than in the 
fermionic model. In the case of the bosonic constituents we must be very 
careful in choosing the binding interaction; the composite boson mass 
approaches zero in the limit of strong coupling $G\rightarrow \infty$ 
\underline{if} we multiply the naive current-current interaction with the factor 
$1/[1+(2G/N)\phi^{\dagger}\phi]$. This factor conspires with $j_{\mu}j^{\mu}$ 
and generates part of the gauge interaction 
$g^2\phi^{\dagger}\phi A_{\mu}A^{\mu}$ with the correct strength 
when we move to the effective theory in terms of the composite field $A_{\mu}$.  
The equation of motion for $A_{\mu}$ prior to generation of the kinetic energy 
is of the form,
\begin{equation}
  A_{\mu} = \frac{gj_{\mu}}{m^2+2g^2\phi^{\dagger}\phi}. \label{NC}
\end{equation} 
where the right-hand side is proportional to the Noether current. Expanding the 
denominator $1/[m^2+2g^2\phi^{\dagger}\phi]$ in the power series of $\phi^{\dagger}\phi$, 
we interpret the series as the composite vector boson consisting not only of 
a single pair of $\phi^{\dagger}\phi$ in $p$ wave but also of many additional 
$\phi^{\dagger}\phi$ pairs in $s$ wave. Even with this additional factor 
$1/[1+2(G/N)\phi^{\dagger}\phi]$, the original Lagrangian $L(\phi^{\dagger},\phi)$ 
written in the $\phi/\phi^{\dagger}$ fields is not gauge invariant at all.  After 
we introduce the composite vector field $A_{\mu}$, however, gauge noninvariance 
is swept entirely into its mass term and we reach the correct form of the gauge 
boson Lagrangian (up to mass).   

Before we proceed to the non-Abelian case, we summarize what we have 
learned from the models that we have so far studied.

1.  In order to generate composite vector bosons with small mass, their
proper self-energy part $\Pi_{\mu\nu}(q)$ must be of the transverse form 
$(-g_{\mu\nu}+q_{\mu}q_{\nu}/q^2)\Pi(q^2)$ without an additional term 
$-g_{\mu\nu}\Pi_0(q^2)$ ($\Pi_0(0)\neq 0)$ since this transverse form
guarantees that the composite mass square is inversely proportional to 
strength of binding interaction.
  
    This transversality is realized in our models with the conserved currents 
of a global symmetry. For this reason, presence of a global symmetry is a 
prerequisite for a generation of (approximate) gauge symmetry though it may 
not be surprising. If conserved currents of a global symmetry do not exist, 
the composite boson mass cannot be made small. By turning the argument around, 
we may say that \underline{if} a tightly bound state of $J^P=1^-$ exists, there must 
be some dynamical reason why such tight binding occurs. Without a good reason
the bound-state mass would only be some fraction of $2M$.  In the $J^P=1^-$
channel a very strong current-current interaction of right properties can
generate a tightly bound state of mass scale much lower than $2M$, at least
theoretically, as we have seen above.

2.  Keeping this observation in mind, we should set up a model Lagrangian 
possessing a global symmetry and introduce nonpropagating composite 
vector-boson fields that are proportional to the Noether currents. 
Then the resulting Lagrangian of the composite fields is gauge invariant 
except for the mass term.  When the Lagrangian is written in the constituent 
particle fields alone, gauge symmetry does not exist since it is broken 
by their kinetic energy and, in the case of boson constituents, by the 
binding interaction too. However, after composite vector bosons are 
generated dynamically, the gauge symmetry breaking is entirely absorbed 
into the boson mass term.   

\section{Case of bosonic constituents: Nonabelian symmetry}

After we have gone through the functional integral formulation of the 
Abelian model, it is not difficult to extend the results to the non-Abelian 
case.  We choose here the bosonic constituents that transform like the 
fundamental representation under flavor SU(n) symmetry. The bosonic 
fields $\phi$ that carry $n$ flavors are replicated with $N$ families as 
$\phi^{(i)}$ ($i=1,2,3\cdots N$). As we have learned in the Abelian case, 
the simplest current-current interaction $(G/N)j_{a\mu}j^{\mu}_a$ summed 
over flavors $a (= 1,2,3\cdots n^2-1)$ does not serve our purpose since 
$j_{a\mu}$ are not conserved currents in the presence of the bosonic 
current-current interaction.  We need the non-Abelian version of the factor 
$1/[1+2(G/N)\phi^{\dagger}\phi$] given in Eq. (\ref{Lb}). The right factor 
in the non-Abelian case is an $(n^2-1)\times(n^2-1)$ matrix in the flavor 
space. It is expressed as the inverse matrix of
\begin{equation}
   (1+\Delta)_{ab} \equiv \delta_{ab} + 
  \biggl(\frac{G}{N}\biggr)\phi^{\dagger}\Bigl\{\frac{\lambda_a}{2},
          \frac{\lambda_b}{2}\Bigr\}\phi,
\end{equation}
where $\frac{\lambda_a}{2}$ are the $n$-dimensional
representation matrices of SU(n). The matrix $(1+\Delta)$ is symmetric
in the $(a,b)$ of flavors and independent of families. In the special case of 
SU(2), $(1+\Delta)$ turns out to be a diagonal matrix in flavors thanks 
to $\{\tau_a,\tau_b\} = 2\delta_{ab}$.  For notational simplicity, 
we suppress hereafter the family indices and often even flavor indices 
when they are obvious. Our Lagrangian is chosen as
\begin{equation}
    L(\phi^{\dagger},\phi) = \partial_{\mu}\phi^{\dagger}\partial^{\mu}\phi
          - M^2\phi^{\dagger}\phi 
        -\biggl(\frac{G}{2N}\biggr)
        j_{a\mu}\Bigl(\frac{1}{1+\Delta}\Bigr)_{ab}j_{b}^{\mu} , 
                       \label{Lbn}
\end{equation}
where summation is understood over flavors $(a,b)$ ($1,2,3\cdots n^2-1$) 
in the interaction term, and $j_{a\mu}$ are the ``naive'' currents 
defined by
\begin{equation}
   j_{a\mu} = i\sum_{i=1}^N \phi^{\dagger(i)}(\lambda_a/2)
          \stackrel{\leftrightarrow}{\partial}_{\mu}\phi^{(i)}.   
\end{equation}  
The flavor and family indices are suppressed altogether in the kinetic 
energy and mass terms. Between the currents $j_{a\mu}$ and $j_b^{\mu}$ is 
the $(ab)$ element of the inverse matrix of $(1+\Delta)$, not the inverse 
of $(1+\Delta)_{ab}$. This is the right current-current interaction that
generates tightly bound vector-boson states of SU(n).
 
The Noether currents of SU(n) can be computed with Eq. (\ref{Lbn}) by using 
the non-Abelian version of Eq. (\ref{jN}) as
\begin{equation}
   J_{a\mu} = \biggl(\frac{1}{1+\Delta}\biggr)_{ab} j_{b\mu}. \label{jNA}
\end{equation}
Being the Noether currents, $J_{a\mu}$ satisfy the conservation law, 
$\partial_{\mu}J_a^{\mu} =0$. By choosing the auxiliary composite fields 
$A_{a\mu}$ proportional to the Noether currents $J_{a\mu}$, we implement
$\partial_{\mu}A_a^{\mu}=0$.   In order to accomplish it, we add
$L_{\rm aux}({\bf A}_{\mu},\phi^{\dagger},\phi)$ to $L(\phi^{\dagger},\phi)$ as
\begin{equation}
 L_{\rm eff}({\bf A}_{\mu},\phi^{\dagger},\phi) = L(\phi^{\dagger},\phi) 
           + L_{\rm aux}({\bf A}_{\mu},\phi^{\dagger},\phi), \label{Lnbeff}
\end{equation}
where 
\begin{eqnarray} 
 L_{\rm aux}({\bf A}_{\mu},\phi^{\dagger},\phi) &=& \frac{1}{2}m^2
   \tilde{A}_{a\mu}(1+\Delta)_{ab}\tilde{A}_b^{\mu} 
    +2(n^2-1)\delta^4(0)\ln{\rm det}(1+\Delta),     \nonumber \\
  \tilde{A}_{a\mu} &=& A_{a\mu} -
              \sqrt{G/Nm^2}\biggl(\frac{1}{1+\Delta}\biggr)_{ab}j_{b\mu}. \label{LnbA}
\end{eqnarray}
The determinant of $(1+\Delta)$ here means the determinant in the flavor SU(n) space. 
It compensates the same term of the 
opposite sign that arises upon the functional integration $\int{\cal D}{\bf A}_{\mu}$ 
in the partition function,
\begin{equation}
  Z = \int{\cal D}{\bf A}_{\mu}{\cal D}\phi^{\dagger}{\cal D}\phi
         \exp\int L_{\rm eff}({\bf A}_{\mu},\phi^{\dagger},\phi) d^4x,
\end{equation}
so that $Z$ is equal to what we have before introducing the fields 
${\bf A}_{\mu}$.  Just as in the Abelian case, the logarithmic term does not 
contribute to our calculation of bound states in the leading $N$ order.

Let us examine the Lagrangian $L_{\rm eff}({\bf A}_{\mu},\phi^{\dagger},\phi)$  
of Eq. (\ref{Lnbeff}). Opening up the mass term of $\tilde{A}_{a\mu}$ and adding 
it to $L(\phi^{\dagger},\phi)$, we find the simple form,
\begin{eqnarray}
  L_{\rm eff}(A_{\mu},\phi^{\dagger},\phi) &=& 
       \partial_{\mu}\phi^{\dagger}\partial^{\mu}\phi
     -  M^2\phi^{\dagger}\phi +2(n^2-1)\delta^4(0)\ln{\rm det}(1+\Delta) \nonumber \\
        &+& \frac{1}{2}m^2A_{\mu}(1+\Delta)A^{\mu} -\sqrt{Gm^2/N}j_{\mu}A^{\mu}, 
                                          \label{Leff}           
\end{eqnarray}
where the flavor indices have been suppressed; the term $j_{\mu}A^{\mu}$ in the
last term stands for $\sum_a j_{a\mu}A_a^{\mu}$. The current-current interaction 
in $L(\phi^{\dagger},\phi)$ is cancelled out by the term arising from 
$\frac{1}{2}m^2\tilde{A}_{\mu}\tilde{A}^{\mu}$ in   
$L_{\rm eff}({\bf A}_{\mu},\phi^{\dagger},\phi)$.

The coefficient $\sqrt{Gm^2/N}$ of $j_{\mu}A^{\mu}$ is identified with the gauge
coupling $g$ in Eq. (\ref{Leff});
\begin{equation}
     g = \sqrt{Gm^2/N}.
\end{equation}
With this relation in mind, we can write the part 
$\frac{1}{2}m^2A_{\mu}\Delta A^{\mu}$ of the term
$\frac{1}{2}m^2A_{\mu}(1+\Delta)A^{\mu}$ in Eq. (\ref{Leff}) explicitly in the
form
\begin{equation}
   \frac{1}{2}m^2A_{\mu}\Delta A^{\mu} = 
    \frac{1}{2}g^2 A_{a\mu}\Bigl(\phi^{\dagger}
   \biggl\{\frac{\lambda_a}{2},\frac{\lambda_b}{2}\biggr\}\phi\Bigr)A^{b\mu}.     
\end{equation}
Therefore the three terms, $\partial_{\mu}\phi^{\dagger}\partial^{\mu}\phi$,
$-gj_{\mu}A^{\mu}$, and $\frac{1}{2}m^2A_{\mu}\Delta A^{\mu}$, add up in the 
gauge-invariant kinetic energy term;
\begin{equation}
     \partial_{\mu}\phi^{\dagger}\partial^{\mu}\phi
     -gj_{\mu}A^{\mu} + \frac{1}{2}m^2A_{\mu}\Delta A^{\mu} = 
 \phi^{\dagger}\biggl(\stackrel{\leftarrow}{\partial}_{\mu}
        -ig\frac{\lambda_a}{2}A_{a\mu}\biggr)
       \biggl(\partial^{\mu}+ig\frac{\lambda_b}{2}A^{\mu}_b\biggr)\phi.
\end{equation}

The remaining task is to generate the kinetic energy term of ${\bf A}_{\mu}$.  
We compute the non-Abelian counterpart of the loop and the bubble in Fig.4 
for the two-point function and, in addition, the three-point and four-point 
functions in the leading $N$ order.  Since $L_{\rm eff}$ is gauge-invariant up 
to the mass of $A_{a\mu}$ and no composite boson loop enters in the leading order, 
this calculation inevitably generates the gauge-invariant combination of 
${\bf A}_{\mu}{\bf A}^{\mu}$, ${\bf A}_{\mu}{\bf A}_{\nu}{\bf A}_{\lambda}$ and 
${\bf A}_{\mu}{\bf A}_{\nu}{\bf A}_{\lambda}{\bf A}_{\kappa}$ as 
\begin{equation}
     -\frac{Z_3}{2}{\rm tr}{\bf G}_{\mu\nu}{\bf G}^{\mu\nu},
\end{equation}          
where ${\bf G}_{\mu\nu}= \frac{1}{2}\lambda_aG_{a\mu\nu}$ is the covariant 
field tensors in matrix,
\begin{equation}
      G_{a\mu\nu} = \partial_{\mu}A_{a\nu}-\partial_{\nu}A_{a\mu}
                     -gf_{abc}A_{b\mu}A_{c\nu}.    
\end{equation}    
Since we know that the final result should come out to be proportional to 
$-\frac{1}{2}{\rm tr}{\bf G}_{\mu\nu}{\bf G}^{\mu\nu}$, we have only to 
compute one of its terms, say, the two-point function.
We find through an explicit diagram calculation
\begin{equation}
     Z_3 = \frac{g^2}{96\pi^2}\ln\frac{\overline{\Lambda}^2}{M^2}.
\end{equation}
As before, the constant $Z_3$ is renormalized away by
\begin{eqnarray}
     {\bf A}_{\mu} &\rightarrow& {\bf A}_{\mu}/\sqrt{Z_3}, \nonumber \\
      g &\rightarrow& \sqrt{Z_3}g \nonumber \\
      m^2 &\rightarrow& Z_3 m^2. \label{renormN}
\end{eqnarray} 
Therefore the final Lagrangian is
\begin{eqnarray}
 L({\bf A}_{\mu},\phi^{\dagger},\phi)  &=&
  -\frac{1}{2}{\rm tr}{\bf G}_{\mu\nu}{\bf G}^{\mu\nu}
  + m^2{\rm tr}{\bf A}_{\mu}{\bf A}^{\nu} \nonumber \\  
    &+& \phi^{\dagger}\biggl(\stackrel{\leftarrow}{\partial}_{\mu}
        -ig\frac{\lambda_a}{2}A_{a\mu}\biggr)
       \biggl(\partial^{\mu}+ig\frac{\lambda_b}{2}A^{\mu}_b\biggr)\phi
        -M^2\phi^{\dagger}\phi    \nonumber \\
    &+& 2(n^2-1)\delta^4(0)\ln{\rm det}(1+\Delta), 
\end{eqnarray}
where it is understood that renormalization has been made for $g$ and $m^2$ as
in Eq. (\ref{renormN}).
This completes our derivation in the non-Abelian case. Clever use of the 
functional integral method streamlines the whole derivation and greatly 
alleviates the calculation that would be quite cumbersome in the diagrammatic 
method.  

A final remark is again on distinction of our study from the 
$CP^{N-1}$ model.  In the strong coupling limit 
of $G\to\infty$ at which the composite boson mass goes to zero, the nonabelian 
Lagrangian in terms of $\phi/\phi^{\dagger}$ approaches the form
\begin{equation}
    \lim_{G\to\infty}L(\phi^{\dagger},\phi) = 
          \partial_{\mu}\phi^{\dagger}\partial^{\mu}\phi - M^2\phi^{\dagger}\phi 
        - 2j_{a\mu}[1/(\phi^{\dagger}D\phi)]_{ab}j_{b}^{\mu}, \label{LlimN}
\end{equation}
where $D_{ab} = \{\lambda_a,\lambda_b\}$. It has no resemblance to the $CP^{N-1}$
model in any respect. In this limit the composite bosons of SU(n) turn massless 
and the Lagrangian becomes gauge invariant under the flavor SU(n). 
The Noether currents, Eq. (\ref{jNA}), vanish at $G\to\infty$ so that there is 
no conflict with the no-go theorem\cite{Case}.  However, it is questionable 
whether such ``gauge bosons'' have any physical significance or even 
exist at all.  (See the remark made on this limit in the Abelian case.)
As for the large $N$ expansion, our large $N$ is the number of families not of 
flavors while it is the number of our flavors $n$ that is made large in the 
computation of the $CP^{N-1}$ model.

\section{Extended model of fermionic constituents}
 
  The condition of tight binding imposes strong constraints on the binding 
interaction.  In fact, it determines the form of interaction almost uniquely.  
After we have gone through our models, we are able to extend the original 
fermionic model a little by adding the $^3D_1$ force proportional 
to $(\overline{\psi}\stackrel{\leftrightarrow}{\partial}_{\mu}\psi)
(\overline{\psi}\stackrel{\leftrightarrow}{\partial}^{\mu}\!\psi)$ to 
the $^3S_1$ force. Let us discuss briefly such a model of abelian symmetry. 
We study the Lagrangian defined by
\begin{eqnarray}
  L(\overline{\psi},\psi) &=&\overline{\psi}(i/\!\!\!\partial-M)\psi \nonumber \\
    &-& \biggl(\frac{G}{2N}\biggr)\frac{(j_{\mu}+(G/N)(\lambda/M^2)
      (\overline{\psi}{\psi})s_{\mu})(j^{\mu}+(G/N)(\lambda/M^2)
      (\overline{\psi}{\psi})s^{\mu})}{1+2(G/N)^2(\lambda/M^2)
       (\overline{\psi}\psi)^2}  \nonumber \\
   &+& \biggl(\frac{G}{4N}\biggr)
     \biggl(\frac{\lambda}{M^2}\biggr)s_{\mu}s^{\mu},
  \;\; (\lambda > 0), \label{LD}                        
\end{eqnarray}
where
\begin{equation}
    j_{\mu}= \overline{\psi}\gamma_{\mu}\psi, \;\;\;
    s_{\mu}= i\overline{\psi}\stackrel{\leftrightarrow}{\partial}_{\mu}\psi.
\end{equation}
Summation over families ($1\sim N$) is understood in $\overline{\psi}\psi$, 
$j_{\mu}$ and $s_{\mu}$, while the flavor of $\psi$ is a simple Abelian charge.  
In Eq. (\ref{LD}) $\lambda$ is a free dimensionless parameter that determines 
the amount of $D$-wave mixing. The value of $\lambda$ must be positive in order 
for the well-defined vacuum to exist. The interaction has been so chosen that 
the Noether current comes out in a reasonably simple form:
\begin{equation}
   J_{\mu} = \frac{j_{\mu} + (G/N)(\lambda/M^2)(\overline{\psi}{\psi})s_{\mu}  
         }{1+2(G/N)^2(\lambda/M^2)(\overline{\psi}{\psi})^2}. \label{NJD}
\end{equation} 
Although the term $s_{\mu}s^{\mu}$ has been introduced to generate the $^3D_1$ 
force with the same order in strength as the $^3S_1$ force, the $^3D_1$ current 
$s_{\mu}$ enters the Noether current $J_{\mu}$ by one power higher in $(G/N)$ 
than the $^3S_1$ current since $j_{\mu}$ arises from the kinetic energy term 
too. Following the procedure in the previous models, we introduce the 
auxiliary field $A_{\mu}$ with the Lagrangian term
\begin{eqnarray}
    L_{\rm aux}(A_{\mu},\overline{\psi},\psi) 
       &=&  (m^2/2)\Bigl(1+2(G/N)^2(\lambda/M^2)(\overline{\psi}\psi)^2\Bigr)
            \tilde{A}_{\mu}\tilde{A}^{\mu} \nonumber \\
   &+& 2\delta^4(0)\ln\Bigl(1+2(G/N)^2(\lambda/M^2)(\overline{\psi}\psi)^2\Bigr),
\end{eqnarray}
where
\begin{equation}
  \tilde{A}_{\mu} = A_{\mu}-\sqrt{\frac{G}{Nm^2}}
          \frac{j_{\mu}+(G/N)(\lambda/M^2)(\overline{\psi}\psi)s_{\mu}
          }{1+2(G/N)^2(\lambda/M^2)(\overline{\psi}\psi)^2},   
\end{equation} 
The Lagrangian $L_{\rm aux}$ leads to the equation of motion for 
$A_{\mu}$, 
\begin{equation}
              A_{\mu} = \sqrt{G/Nm^2}J_{\mu}. 
\end{equation} 
The field $A_{\mu}$ obeys $\partial_{\mu}A^{\mu}=0$ and the proper 
self-energy part is transverse.  Adding $L_{\rm aux}$ to 
$L(\overline{\psi},\psi)$, we have
\begin{eqnarray}
  L_{\rm eff}(A_{\mu},\overline{\psi},\psi) &=& (m^2/2)A_{\mu}A^{\mu} +
      \overline{\psi}(i/\!\!\!\partial-M)\psi - gj_{\mu}A^{\mu}
                                \nonumber \\   
     &+&\biggl(\frac{\lambda G}{4NM^2}\biggr)\Bigl(s_{\mu}s^{\mu}  
       - 4g(\overline{\psi}\psi)s_{\mu}A^{\mu}  
     + 4g^2(\overline{\psi}\psi)^2A_{\mu}A^{\mu}\Bigr) \nonumber \\
 &+& 2\delta^4(0)\ln\Bigl(1+2(G/N)^2(\lambda/M^2)(\overline{\psi}\psi)^2\Bigr),     
\end{eqnarray}
where $g\equiv m\sqrt{G/N}$.
Note that the field $A_{\mu}$ enters $L_{\rm eff}$ precisely in the 
gauge-invariant form up to the mass term.  Therefore, upon generating the kinetic 
energy of $A_{\mu}$ by loops and bubbles and renormalizing $Z_3$ away by
$A_{\mu}\to A_{\mu}/\sqrt{Z_3}$, $g\to g/\sqrt{Z_3}$ and $m^2\to m^2/Z_3$, 
we reach in terms of the renormalized mass and coupling
\begin{eqnarray}
  L(A_{\mu},\overline{\psi},\psi) &=& -\frac{1}{4}F_{\mu\nu}F^{\mu\nu}
              + \frac{m^2}{2}A_{\mu}A^{\mu}  \nonumber \\
    &+& \overline{\psi}(i/\!\!\!\!D-M)\psi + \frac{\lambda G}{4NM^2}
        (i\overline{\psi}\stackrel{\leftrightarrow}{D}_{\mu}\psi)
                      (i\overline{\psi}\stackrel{\leftrightarrow}{D}^{\mu}\psi)
                                      \nonumber \\
  &+& 2\delta^4(0)\ln\Bigl(1+2(G/N)^2(\lambda/M^2)(\overline{\psi}\psi)^2\Bigr),  
                     \label{LDA}
\end{eqnarray}
where $F_{\mu\nu}=\partial_{\mu}A_{\nu}-\partial_{\nu}A_{\mu}$, $D_{\mu}=
\partial_{\mu}+igA_{\mu}$.  The $^3D_1$ four-fermion interaction does not 
go away in Eq. (\ref{LDA}) but gauged with $A_{\mu}$. The sole 
gauge-noninvariant term is the mass term $\frac{1}{2}m^2A_{\mu}A^{\mu}$. 
Although it looks tempting to introduce another auxiliary vector field 
$B_{\mu}$ to remove the ``gauged $s_{\mu}s^{\mu}$ term'' from the Lagrangian 
of Eq. (\ref{LDA}), it is not possible since the coefficient of 
$s_{\mu}s^{\mu}$ is positive (repulsive).\footnote{Positivity of $\lambda$ 
is required by existence of a well-defined classical vacuum.  For $\lambda <0$ 
the denominator $1/[1+2(G/N)^2(\lambda/M^2)(\overline{\psi}\psi)^2]$ would 
blow up at $(\overline{\psi}\psi)=\sqrt{-1/2\lambda}(N/G)M$ in Eq. (\ref{LD}).}

   Since the $^3D_1$ four-fermion interaction stays in the Lagrangian 
$L_{\rm eff}(A_{\mu},\overline{\psi},\psi)$, the wave-function renormalization 
$Z_3$ and therefore the mass $m^2$ and the coupling $g^2$ are to be computed 
in the perturbation series with respect to $\lambda$. We rewrite $G/N$ in terms 
of $g$ and $m^2$ by use of $G/N=g^2/m^2$ and carry out the computation.  
In the zeroth order of $\lambda$ the simple 
one-loop-diagram of fermion is transverse by itself and generates 
$Z_3 = (g^2N/12\pi^2)\ln(\overline{\Lambda}^2/M^2)$. In the first order of 
$\lambda$ there exist four diagrams which sum up to the transverse form.  
(See. Fig. 5.) When we compute divergent integrals by the dimensional 
regularization, we find that the $O(\lambda)$ correction to $Z_3$ happens to 
vanish by cancellation among the four diagrams. 

\noindent
\begin{figure}[h]
\epsfig{file=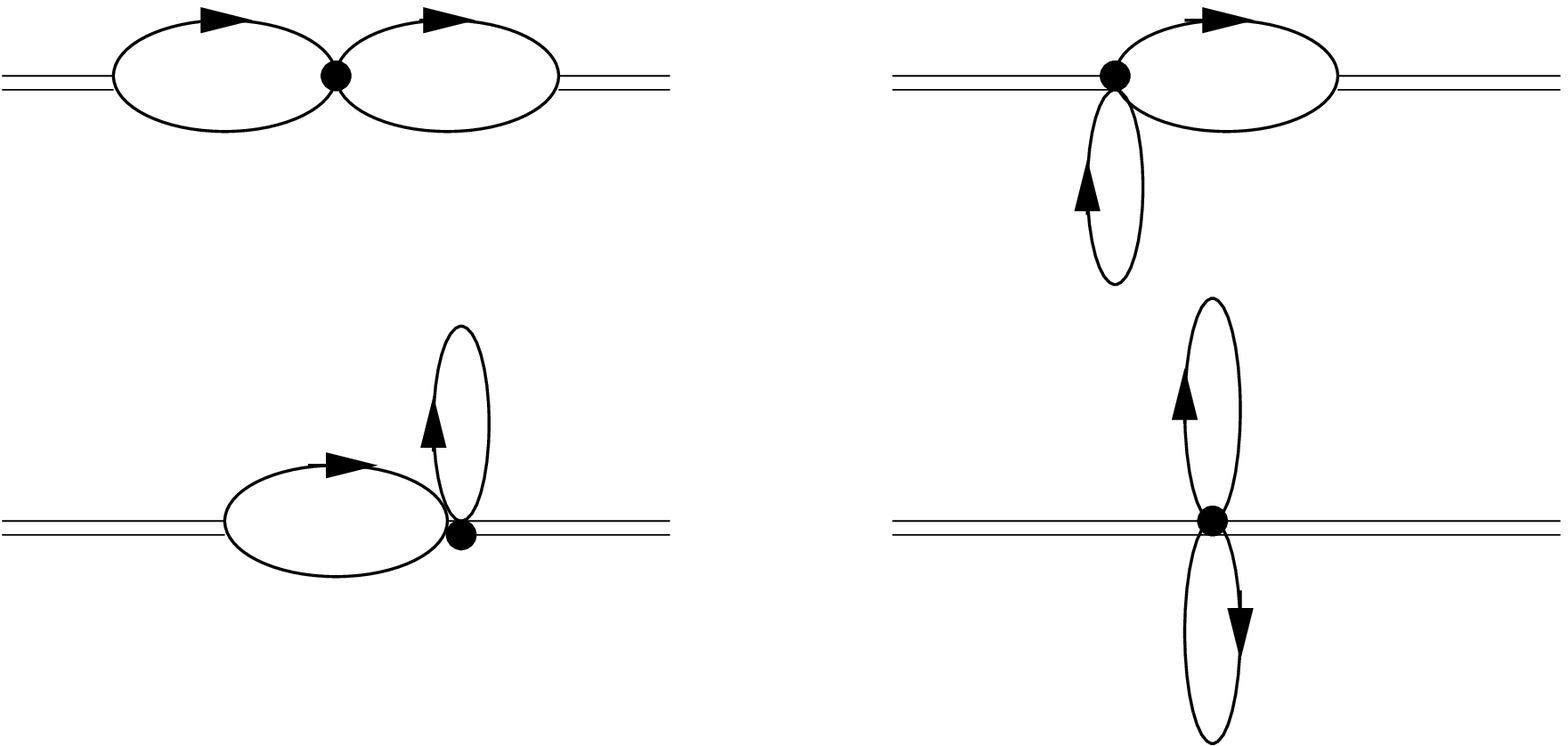,width=8cm,height=4cm} 
\caption{The self-energy diagrams of $O(\lambda)$. The filled small circles
denote the locations where the interaction proportional to $\lambda$ enters.        
\label{fig:5}}
\end{figure}

In the order of $\lambda^2$ there are again four diagrams, which differ from 
the diagrams of $O(\lambda)$ by insertion of a single fermion loop 
$\int e^{iqx}\langle 0|T(s_{\mu}(x)s_{\nu}(0))|0\rangle d^4x$. 
This insertion maintains the cancellation that occurs among the four diagrams
of $O(\lambda)$. The same cancellation repeats to all higher orders of $\lambda$. 
Consequently we have no $\lambda$ correction to $Z_3$:
\begin{equation}
    Z_3 = (g^2N/12\pi^2)\ln(\overline{\Lambda}^2/M^2)
\end{equation}
to all orders of $\lambda$ in the large $N$ expansion. Therefore the renormalized 
mass and coupling are given by $m^2=12\pi^2/G\ln(\overline{\Lambda}^2/M^2)$ and 
$g^2=12\pi^2/N\ln(\overline{\Lambda}^2/M^2)$.  Absence of the $\lambda$ 
correction is unexpected.  We are unable to appreciate if it has an important
implication or not.

This extended fermionic model reinforces the claim that the generation mechanism 
of approximate gauge symmetry is not an accident but more a general phenomenon.  
As we have emphasized repeatedly, however, it would be a futile effort to try 
to improve the Lagrangian further so as to generate genuine gauge bosons of zero 
mass as composite states unless some local symmetry is slipped in. It is because it 
would contradict with the simple general theorem\cite{Case}. In our case
the Lorentz-covariant 
conserved currents with nonzero charge do exist in the composite vector-boson 
theories as we can write them in terms of the constituent particle fields. 
We have shown for each model in this paper that the conserved (Noether) currents 
would disappear and the fundamental Lagrangian would become meaningless when 
one took the massless limit. In the extended fermion model the Noether current
would disappear and the Lagrangian would become singular at 
$\overline{\psi}\psi = 0$. 

\section{Discussion and outlook}

 Gauge symmetries in particle physics are broken symmetries except for 
electrodynamics and chromodynamics. The prevailing wisdom for broken 
gauge symmetries is that they are spontaneously broken since otherwise 
the underlying quantum field theory would be unrenormalizable.  If we want 
to construct an ultimate fundamental theory valid at all possible energies 
from top down, postulating gauge symmetry is the only option for us. 
On the other hand there may still be layers of effective theories before 
we reach the ultimate theory at the highest energy. Indeed this was the case 
in the history of phenomenological particle physics. If one takes this viewpoint, 
one may rather build particle theory from the bottom up with effective theories which 
are valid only over limited ranges of the energy scale.  It would not be so 
unreasonable for theorists in this camp to ask whether there is any dynamical 
origin for approximate gauge symmetries other than spontaneous symmetry breaking. 

    The purpose of this paper is to show in the solvable models that even 
if a gauge symmetry is not implanted at a fundamental level, it may emerge 
as an approximate symmetry by dynamical necessity in the tightly bound 
limit of composite vector bosons if such bosons exist at all. We know that the 
Nambu-Goldstone boson can be a tightly bound composite massless boson: It appears 
upon spontaneous breaking of a global symmetry and a phase transition occuring.  
In our case a global symmetry remains unbroken and no phase transition occurs.
The tightly bound composite boson is not unnatural in the $J^P=1^-$ channel 
when the composite field is proportional to a Lorentz-covariant conserved 
current. In contrast, in other channels one must fine-tune coupling strength 
if one wants to generate a very light but nonzero composite boson.  We have 
postulated a global symmetry as our starting point to derive an approximate 
local symmetry. Some may ask why we accept a global symmetry at the beginning. 
Are global symmetries more natural than local symmetries ?  Frankly, we cannot 
make a convincing argument in this regard. 

In low-energy strong interactions of mesons and baryons the relation between 
vector mesons and conserved hadronic currents was emphasized by 
Sakurai\cite{Sakurai} nearly a half century ago. He strongly advocated that
the $\rho$, the $\omega$, and the $\phi$ meson couple to the isospin, 
the baryonic, and the hypercharge current, respectively, in the form
$L_{\rm int} = - gj_{\mu}\varphi^{\mu}$ incorporating the $\omega-\phi$ mixing.  
A decade later, from a field theory standpoint, Kroll, Lee and Zumino\cite{Lee} 
proposed the field-current identity hypothesis $\varphi_{\mu} = fj_{\mu}$ in 
which the fields of the $\rho$, the $\omega$ and the $\phi$ meson are in 
fact the isospin, the baryon and the hypercharge current themselves. Our 
finding in this paper reminds us of this old hypothesis although in the 
contemporary picture those light vector mesons are the loosely bound states by 
the long-distance confining forces. Nonetheless, these hypotheses on the light 
vector mesons were successfully tested, for instance, in the vector-meson 
dominance of the electromagnetic and weak currents albeit within the accuracy 
of typical low-energy strong interaction physics. Many years later, but before 
high-energy electroweak interaction data were accumulated, Claudson, Farhi 
and Jaffe \cite{Jaffe} 
proposed that $W$ and $Z$ might be \underline{loosely} bound composite 
bosons by some hypothetical confining force. Criticism was made by Lee and 
Shrock \cite{Shrock} with lattice gauge theory analysis. Beyond that, however, 
conspicuously missing was a quantitative study. The idea of the loosely 
bound $W$ and $Z$ would have hard time to withstand test of the contemporary 
experimental data with respect to the fast falling form-factor damping, 
e.g., large difference between the on-shell coupling and the 
zero-momentum limit of coupling.  More recently, however, attempts have
been made for composite $W$ and $Z$ with higher confinement energy scales 
involving the extra space-time dimension\cite{Gher}. The guage symmetry is
placed at onset outside the four-dimensional spacetime in those models.

  Our field theory models here are all based on unrenormalizable field 
theories in the large $N$ limit since otherwise we cannot solve them explicitly.  
When the models are written in the effective Lagrangian of the composite 
vector bosons, unrenormalizability is transformed into the {\em longitudinal} 
polarization of the massive vector bosons and, in the presence of derivative
interactions, possibly the nonderivative gauge-invariant logarithmic term. 
In this sense our ignorance in the binding interactions is swept into the 
longitudinal polarization state of the composite vector-boson. As it is well 
known\cite{Quigg}, the {\em tree diagrams} involving the self-coupling of 
longitudinal polarizations of the $W$ and $Z$ bosons overshoot the unitarity 
bound at energies much higher than the $W$ and $Z$ masses when the Higgs 
boson is left out or very heavy ($>$ 1 TeV) in the standard model. A possibility 
of building an alternative to the standard model with composite $W$ and $Z$ 
was suggested\cite{Georgi} by introducing a set of sufficiently many new 
fermions as their consituents in our simplest fermionic model. In such models 
$W$ and $Z$ would interact strongly at very high energies through the 
longitudinal polarization modes. This alone does not rule out the composite $W$ 
and $Z$ at present.  However, there exists a potential problem of the same 
origin at lower energies. That is, the radiative corrections to the low-energy 
electroweak parameters.  We can examine the composite vector-boson propagator 
with the diagram of Fig. 1a by taking the external fermion lines off mass shell. 
It is given by
\begin{equation}
         D_{\mu\nu}(q) = i\frac{g_{\mu\nu}-(q_{\mu}q_{\nu}/m^2)F(q^2)}{
         m^2-q^2 F(q^2)}, 
\end{equation}
where the form factor $F(q^2)$ is defined with
$\overline{\Pi}(q^2)\equiv\Pi(q^2)/q^2$ by
\begin{equation}
     F(q^2) = \overline{\Pi}(q^2)/\overline{\Pi}(0), 
\end{equation}  
so that $F(q^2)$ is normalized as $F(0) = 1$. The function $D_{\mu\nu}(q)$ does 
not deviate much from that of the lowest-order perturbation in the region of 
$q^2 = O(m^2) \ll O(M^2)$.\footnote{ 
The damping effect of $F(q^2)$ is measured by its radius defined by 
$\sqrt{\langle r^2\rangle}$ where $dF(q^2)/dq^2|_{q^2=0}=\frac{1}{6}
\langle r^2\rangle$. In our fermionic model $\langle r^2\rangle = 
6/[5M^2\ln(\overline\Lambda^2/M^2)]$, which is $O(1/M^2)$ as we expect.}
Since models of composite $W$ and $Z$ do not contain the Higgs bosons, 
the mass singularity term $q_{\mu}q_{\nu}/m^2$ potentially generates 
large radiative corrections\footnote{
It was argued years ago\cite{R} that for some four-fermion interaction 
theory may become renormalizable when it is written in terms of 
collective modes, {\em i.e.,} composite fields. It does not seem to 
happen in our case of vector bosons.} 
to the low-energy parameters, particularly 
in the $S$ parameter. However, the diagrams which contain a composite 
boson loop are in the next-to-leading order of the large $N$ expansion. 
That is, it is technically outside of our scope of calculation. 
Nonetheless it may become a problem if we seriously attempt to build 
a model of composite $W$ and $Z$ as an alternative to the standard model. 

We all agree that despite its field theoretical beauty the standard model 
has disturbing unnaturalness, the worst of it being the hierarchy problem, 
once we go beyond the multi-TeV energy scale. We should not completely 
abandon esoteric possibilities such as composite $W$ and $Z$ at some very 
high energy-scale until an experiment rules them out convincingly. We should 
keep our mind open for the outcome of the upcoming accelerator experiment 
although admittedly chances may be small. Even if the LHC does not support 
the composite $W$ and $Z$ bosons, it may discover novel spin-one bosons
that interact like gauge bosons. Aside from an experiment, the quest for the 
origin of gauge symmetry will remain a challenge for many 
theorists\cite{Nielsen}.

\acknowledgments

The author acknowledges useful conversations with Korkut Bardakci.
This work was supported by the Director, Office of Science, Office of
High Energy and Nuclear Physics, Division of High Energy Physics,
of the U.S.  Department of Energy under Contract No. DE--AC02--05CH11231. 

\appendix
\section{The functional determinant}

Change of the integral variable from $A_{\mu}$ to 
$(1+2(G/N)\phi^{\dagger}\phi)A_{\mu}$ in
the functional integral of Eq. (\ref{AbA}) is not so trivial as that 
in the ordinary integrals. Although the resulting {\rm logarithmic term}
does not contribute to the final results of our particular 
computation, a remark should be made in order to assure that this change 
of variable does not generate a new gauge-symmetry breaking.   

We go to the Euclidean metric by $it \to t$ and $iE \to E$ and examine the 
functional integral  
\begin{equation}
    \int {\cal D}A_{\mu}\exp\Bigl[-\int  
       \frac{m^2}{2}\Bigl(1+2(G/N)\phi^{\dagger}\phi\Bigr)
           \tilde{A}_{\mu}\tilde{A}_{\mu}d^4x\Bigr],   \label{int}
\end{equation}
where $\tilde{A}_{\mu} = A_{\mu}-\sqrt{G/Nm^2}j_{\mu}/[1+2(G/N)
\phi^{\dagger}\phi]$. We may drop the tilde of $\tilde{A}_{\mu}$ 
by shifting the functional space of $A_{\mu}$ by 
$\sqrt{G/Nm^2}j_{\mu}/[1+2(G/N)\phi^{\dagger}\phi]$. The factor 
$\frac{1}{2}m^2$ may also be dropped since the rescaling of $A_{\mu}(x)$ 
by a constant affects only an unphysical constant factor to
the partition function. However, the multiplication of a function on 
$A_{\mu}$ cannot be dropped in general since it deforms the functional 
phase space.  For notational simplicity, we study for one of the four 
space-time components of $A_{\mu}$ suppressing its subscript for a while. 
The integral of our interest is therefore: 
\begin{equation}
    \int {\cal D}A\exp\Bigl[-\int 
       (1+f(x))A(x)A(x)d^4x\Bigr],   \label{inter}
\end{equation}
where $f(x)\equiv 2(G/N)\phi^{\dagger}(x)\phi(x)$.  Expand $A(x)$ 
in a complete set of orthonormal functions 
$\varphi_i(x)$ ($i=1,2,3,\cdots\infty$) in the 4-dimensional space-time as
\begin{equation}
         A(x) = \sum_i a_i\varphi_i(x),
\end{equation} 
where $\int\varphi_i(x)\varphi_j(x)d^4x = \delta_{ij}$. The functional integral 
Eq. (\ref{inter}) turns into
\begin{equation}
      \int\cdots\int \Pi_k da_k \exp[-\sum_{ij}a_i(\delta_{ij}+f_{ij})a_j],
                                                                \label{X}
\end{equation}
where $f_{ij}=\int\varphi_i(x)f(x)\varphi_j(x)d^4x$. If we choose specifically
the complete set with which the matrix $f_{ij}$ is diagonal, the 
integral $\int da_k$ can be carried out with the quadrature integral
formula as
\begin{eqnarray}
   \Pi_k \int da_k\exp[-(1+f_{kk})a_k^2] 
 &=&{\rm const.}\times 1/\Bigl(\Pi_i\sqrt{1+f_{ii}}\Bigr), \nonumber \\
 &=&{\rm const.}\times \exp\Bigl[-\frac{1}{2}\ln\Bigl(\Pi_i(1+f_{ii})\Bigr)\Bigr]. 
                    \label{matrix}
\end{eqnarray}
Since $\Pi_i(1+f_{ii})$ is the determinant of the infinite-dimensional diagonal 
matrix $(1+f_{ii})\delta_{ij}$, the last line of Eq. (\ref{matrix}) can be 
expressed as 
\begin{equation}
     {\rm const.}\times\exp\Bigl(-\frac{1}{2}\ln{\rm det}(1+f^{(D)})\Bigr), \label{C}
\end{equation}
where we have supplied the superscript $D$ to $f$ in order to emphasize that $f^{(D)}$ 
is a diagonal matrix.  The undetermined (infinite) multiplicative constant
in front of the exponent is absorbed into the ill-defined measure of functional 
phase space that has no physical effect.

Going back to Eq. (\ref{matrix}), let us expand the logarithm with the
Taylor series expansion formula of $\ln(1+\xi)$ as 
\begin{eqnarray}
 \ln(\Pi_i(1+f_{ii})) &=& \sum_{i}\ln(1+f_{ii}), \nonumber \\
                     &=& \sum_{n=1}^{\infty}
          \sum_i\Bigl(\frac{(-1)^{n-1}}{n} (f_{ii})^n\Bigr).  \label{trace}
\end{eqnarray}
Note that for the diagonal matrix, $\sum_i (f_{ii})^n = \sum_i (f^n)_{ii}
={\rm tr}(f^n)$ and furthermore that a trace of the matrix element does not 
depend on the choice of its basis. Therefore, we can go to the four-dimensional 
Fourier basis ($i\to k_1,k_2,k_3,k_4$) and rewrite Eq. (\ref{trace}) as
\begin{eqnarray}
 \ln(\Pi_i(1+f_{ii})) &=& \sum_{n=1}^{\infty}\frac{(-1)^{n-1}}{n}
   \int\frac{d^4k}{(2\pi)^4}\Bigl(\int e^{-ikx}f(x)^n e^{ikx} d^4x\Bigr), \nonumber \\
   &=&\biggl(\int\sum_{n=1}^{\infty}\frac{(-1)^{n-1}}{n}f(x)^n d^4x\biggr)\delta^4(0),          
\end{eqnarray} 
where the last factor $\delta^4(0)$ comes from 
\begin{equation}
\int d^4k/(2\pi)^4 = 
  \lim_{y\to z}\int e^{ik(y-z)}d^4k/(2\pi)^4 =\lim_{y\to z}\delta^4(y-z).
\end{equation}
In the diagram calculation the function $\delta^4(0)$ arises from the quartic 
divergence $\Lambda^4/32\pi^2$ of the $A_{\mu}$-bubble diagram, as will be shown 
later in this Appendix.  Putting $m^2$ back in Eq. (\ref{inter}), we reach 
\begin{eqnarray}
\!\!\int\!{\cal D}A_{\mu}\exp\Bigl(-\int
   \frac{m^2}{2}(1+f(x))A_{\mu}A^{\mu}d^4x\Bigr) &=&  
   \exp\Bigl(-2\delta^4(0)\int\sum_{n=1}^{\infty}\frac{(-1)^{n-1}}{n}f(x)d^4x\Bigr),
                          \label{expansion} \\
      &=& \exp\Bigl(-2\delta^4(0)\int\ln(1+f(x))d^4x\Bigr),
\end{eqnarray}
where the four space-time components of $A_{\mu}$ generate four identical 
terms to turn $\frac{1}{2}$ into $4\times \frac{1}{2}\rightarrow 2$ in
the right-hand side of Eq. (\ref{expansion}).  The irrelevant constant 
in front has been suppressed above. \\

\underline{Diagrammatic explanation}

In the remainder of Appendix we show the diagrammatic origin of this 
logarithmic term.  
Let us expand both 
sides of Eq. (\ref{expansion}) in the power series of $G$ and compare order
by order the right-hand side with their corresponding diagrams computed with 
$L_{\rm eff}(A_{\mu}, \phi^{\dagger}, \phi)$ of Eq. (\ref{Lg}) in the left-hand 
side. Our purpose here is pedagogical; We show that the integration over 
$\int{\cal D}A_{\mu}$ generates Green's functions from pairs of $A_{\mu}$ and 
indeed leads to the logarithmic term in Eq. (\ref{expansion}). 
The vector-boson two-point function for the Lagrangian of Eq. (\ref{Lg}) 
is given by
\begin{equation}
  \langle 0|T(A_{\mu}(x)A_{\nu}(y))|0\rangle 
                = (1/m^2)\delta_{\mu\nu}\delta^4(x-y), 
\end{equation}
since there is no kinetic energy term of $A_{\mu}$ at this stage.

The term of $O(G)$ in the right-hand side is 
$-2\delta^4(0)\int(2G/N)\phi^{\dagger}\phi d^4x$. 
This arises from the diagram 
Fig. 6a that consists of a single Green's function of $A_{\mu}$:
\begin{eqnarray}
  O(G)_{\rm left}&=& -\int\frac{m^2}{2}(2G/N)\phi^{\dagger}(x)\phi(x)
   \langle 0|T(A_{\mu}(x)A_{\mu}(x))|0\rangle d^4x\nonumber \\
       &=& -4(G/N)\delta^{4}(0)\int\phi^{\dagger}(x)\phi(x)d^4x. \label{G}
\end{eqnarray} 
This term is the quartically divergent self-energy of $\phi/\phi^{\dagger}$,
but cancelled out in the final answer by the one-loop self-energy diagram 
$O((\sqrt{G/N})^2)$ of the interaction $\sqrt{G/N}j_{\mu}A^{\mu}$.

When we move to the order $O(G^2)$ and higher, there exist the contributions 
of connected and disconnected diagrams. 
The terms of $G^2$ in the expansion of the right-hand side of
Eq. (\ref{expansion}) are
\begin{equation}
  O(G^2)_{\rm right} = (-2\delta^4(0))\times \frac{-1}{2}(2(G/N))^2 \int 
             (\phi^{\dagger}(x)\phi(x))^2 d^4x 
         + \frac{1}{2!}[O(G)]^2,  \label{G2}
\end{equation}
where $[O(G)]^2$ means square of the term of $O(G)$ in Eq. (\ref{G}), that is, 
the disconnected diagram of two $O(G)$ bubbles (the first diagram of Fig. 6b).  
The first term of Eq. (\ref{G2}) comes from the connected diagram 
in Fig. 6b:
\begin{eqnarray}
  O(G^2)_{\rm left\; connected}&=&  
        2\times(-1)^2\times\frac{1}{2!}\Bigl(\frac{1}{2}m^2\Bigr)^2 (2G/N)^2
       \int\;\int (\phi^{\dagger}(x)\phi(x))(\phi^{\dagger}(y)\phi(y))d^4xd^4y
             \nonumber     \\ 
        &\times& \langle 0|T(A_{\mu}(x)A_{\nu}(y))|0\rangle
                \langle 0|T(A_{\mu}(x)A_{\nu}(y))|0\rangle ,  \label{4} \\
      &=&  (2G/N)^2\delta^4(0)\int (\phi^{\dagger}(x)\phi(x))^2 d^4x,  
\end{eqnarray}
where the first factor 2 comes from two different ways of matching $A_{\mu}$ 
fields into two-point functions and $\frac{1}{2!}$ is from the second-order
perturbation expansion. Summation over subscripts $\mu$ and $\nu$ 
generates the factor of 4 in the last line. This agrees with the $(n=2)$ term
of Eq. (\ref{expansion}) in the expansion. The term proportional to 
\begin{equation}
\langle 0|T(A_{\mu}(x)A_{\mu}(x))|0\rangle
                \langle 0|T(A_{\nu}(y)A_{\nu}(y))|0\rangle 
\end{equation}
appears from the disconnected diagrams in the left-hand side and matches
the second-order Taylor expansion of the $(n=1)$ term in the right-hand
side of Eq. (\ref{expansion}).

\noindent
\begin{figure}[h]
\epsfig{file=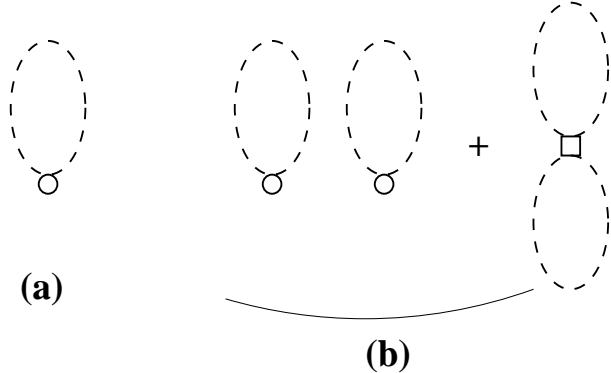,width=8cm,height=5cm} 
\caption{Breakdown of the logarithmic term of the functional integral in
terms of diagrams. (a) $O(G)$ and (b) $O(G^2)$. The small circle
represents the local limit of two-point Green's function of $A_{\mu}$ while 
the small square is for the local limit of the connected
Green's function of four $A_{\mu}$'s (Eq. (\ref{4})), while the ellipse in 
a broken line represents $\phi^{\dagger}(x)\phi(x)$.       
\label{fig:6}}
\end{figure}

We can go on to $O(G^3)$ and higher-order terms. The $n=3$ term in the exponent 
of the right-hand side is 
$-\frac{2}{3}\delta^4(0)\int [2(G/N)\phi^{\dagger}(x)\phi(x)]^3 d^4x$, 
while the connected diagrams of $O(G^3)$ in the left-hand side matches this
term:
\begin{equation}
  O(G^3)_{\rm left\; connected} =
  \frac{1}{3!}\times 8\biggl(-\frac{m^2}{2}\biggr)^3\times
  4\biggl(\frac{1}{m^2}\biggr)^3\delta^4(0)
 \int\Bigl(2(G/N)^3\phi^{\dagger}(x)\phi(x)\Bigr)^3 d^4x,
\end{equation} 
where the factor $1/3!$ comes from the third-order perturbation expansion,
the factor $8$ in front is due to eight ways to pair $A_{\mu}$'s into two-point 
Green's functions, the factor 4 in front of $(1/m^2)^3$ results from the
sum over the polarization subscript of $A_{\mu}$, and each $(1/m^2)$ comes from 
a Green's function of $A_{\mu}$. The disconnected terms match in much the
same way as in the case of $O(G^2)$.



\begin{thebibliography}{}

\bibitem{Heisenberg} W. Heisenberg, Rev. Mod. Phys. {\bf 29}, 269 (1957).\\
H. P. D\"{u}rr, Heisenberg, H. Mitter, S. Schlieder, and K. Yamazaki,
Z. Naturforsh. {\bf 14A}, 441 (1959). \\
J. D. Bjorken, Ann. Phys. (N.Y.) {\bf 24}, 174 (1963). \\
I. Bialynicki-Birula, Phys. Rev. {\bf 130}, 465 (1963). \\
T. Eguchi and H. Sugawara, Phys. Rev. D {\bf 10}, 4257 (1974); T. Eguchi, 
  Phys. Rev. D {\bf 14}, 2755 (1976). \\
K. Kikkawa, Progr. Theor. Phys. {\bf 56}, 947 (1976).  \\
Y. Terazawa, Y. Chikashige, and K. Akama, Phys. Rev D {\bf 15}, 480 (1977); 
K. Akama, Phys. Rev. Lett. {\bf 76}, 184 (1996); K. Akama and T. Hattori,  Phys. 
Lett. B {\bf 393}, 383 (1997). \\
\bibitem{Nielsen} C. D. Froggatt and H. B. Nielsen, {\em Origin of Symmetries}
(World Scientific, Singapore, 1991) and references therein. \\
J. L. Chikaleuli, C. D. Froggatt, and H. B. Nielsen, Phys. Rev. Lett. {\bf 87} 
091601 (2001).
\bibitem{Mandel} M. Veltman, Acta Phys. Pol. {\bf B12}, 437 (1981).\\ S. Mandelstam, 
{A Passion in Physics} edited by C. DeTar {\em et al.} (World Scientific, Singapore, 
1985), p.97. 
\bibitem{Suzuki} M. Suzuki, Phys. Rev. D {\bf 37}, 210 (1988).
\bibitem{BJ} J. D. Bjorken, Phys. Rev. D {\bf 19}, 335 (1979).
\bibitem{HS} P. Q. Hung and J. J. Sakurai, Nucl. Phys. {bf B143}, 81 (1978);
  {\bf 148B}, 538(E) (1979).
\bibitem{Georgi} A. Cohen, H. Georgi, and E. Simmons, Phys. Rev. D {\bf 38}, 405 (1988).
\bibitem{Bando}  M. Bando, T. Kugo, and K. Yamawaki, Phys. Rep. {\bf 164}, 217 (1988).  
\bibitem{Case} K. M. Case and S. Gasiorowicz, Phys. Rev. {\bf 125}, 1055 (1962).\\
S. Weinberg and E. Witten, Phys. Lett. {\bf 96B}, 59 (1980). 
\bibitem{Haber} H. E. Haber, I. Hinchliffe, and E. Rabinovici, Nucl. Phys. {\bf B 172}, 
458 (1980).
\bibitem{Akh} E. K. Akhmedov, Phys. lett. {\bf B521}, 79 (2001).
\bibitem{Sakurai} J. J. Sakurai, Ann. Phys. (N.Y.) {\bf 11}, 1 (1960).
\bibitem{Lee} N. M. Kroll, T. D. Lee, and B. Zumino, Phys. Rev. {\bf 157}, 1376 (1967); 
T. D. Lee and B. Zumino, Phys. Rev. D {\bf 163}, 1667 (1967).
\bibitem{Jaffe}   M. Claudson, E. Farhi, and R. L. Jaffe, Phys. Rev. D {\bf 34}, 873 (1986).\\
\bibitem{Shrock} I-H. Lee and R. E. Shrock, Phys. Rev. Lett. {\bf 59}, 14 (1987); 
 Phys. Lett. {\bf 199B}, 541 (1987); Phys. Lett. {\bf 201B}, 497 (1988); 
S. Aoki, I-H. Lee, and R. E. Shrock, Phys. Lett. {\bf 207B}, 471 (1988). 
\bibitem{Gher} For instance, Y. Cui, T. Gherghetta, and J. D. Wells, J High Energy Phys. 
09 (2009) 080 and references therein.
\bibitem{Quigg} B. W. Lee, C. Quigg, and H. Thacker, Phys. Rev. Lett. {\bf 38}, 883 
(1977); Phys. Rev. D {\bf 16}, 1519 (1977). 
\bibitem{R} See J. D. Bjorken, in [1]. \\ 
I Bialynicki-Birula, Phys. Rev. {\bf 130}, 465 (1963).\\  G. S. Guralnik, Phys. Rev. 
{\bf 136}, B1404 (1964).\\  T. Eguchi, Phys. Rev. D {\bf 17}, 611 (1978).
\end{thebibliography}
\end{document}